\documentclass[twocolumn,tighten]{aastex62}
\pdfoutput=1 
\usepackage{amsmath,amstext}
\usepackage[T1]{fontenc}
\usepackage{apjfonts} 
\usepackage{xcolor}
\usepackage[figure,figure*]{hypcap}

\usepackage{todonotes}

\newcommand*{\hmpc}{h^{-1}\textrm{Mpc}}

\newcommand*{\hgpc}{h^{-1}\textrm{Gpc}}
\newcommand{\hmsun}{h^{-1}\textrm{M}_{\odot}}
\newcommand{\redmapper}{redMaPPer}

\newcommand\aemulus{{\sc Aemulus}}

\shorttitle{Aemulus II: Emulating the Halo Mass Function}
\shortauthors{McClintock, T., et al.}

\begin{document}

\title{The Aemulus Project II: Emulating the Halo Mass Function}

\author{Thomas McClintock}
\affiliation{Department of Physics, University of Arizona, Tuscon, AZ 85721, USA}

\author{Eduardo Rozo}
\affiliation{Department of Physics, University of Arizona, Tuscon, AZ 85721, USA}

\author{Matthew R. Becker}
\affiliation{Kavli Institute for Particle Astrophysics and Cosmology and Department of Physics, Stanford University, Stanford, CA 94305, USA}
\affiliation{Department of Particle Physics and Astrophysics, SLAC National Accelerator Laboratory, Stanford, CA 94305, USA}
\affiliation{Civis Analytics, Chicago, IL 60607, USA}

\author{Joseph DeRose}
\affiliation{Kavli Institute for Particle Astrophysics and Cosmology and Department of Physics, Stanford University, Stanford, CA 94305, USA}
\affiliation{Department of Particle Physics and Astrophysics, SLAC National Accelerator Laboratory, Stanford, CA 94305, USA}

\author{Yao-Yuan Mao}
\affiliation{Department of Physics and Astronomy and the Pittsburgh Particle Physics, Astrophysics and Cosmology Center (PITT PACC), University of Pittsburgh, Pittsburgh, PA 15260, USA}

\author{Sean McLaughlin}
\affiliation{Kavli Institute for Particle Astrophysics and Cosmology and Department of Physics, Stanford University, Stanford, CA 94305, USA}
\affiliation{Department of Particle Physics and Astrophysics, SLAC National Accelerator Laboratory, Stanford, CA 94305, USA}

\author{Jeremy L. Tinker}
\affiliation{Center for Cosmology and Particle Physics, Department of Physics, New York University, 4 Washington Place, New York, NY 10003, USA}
 
\author{Risa H. Wechsler}
\affiliation{Kavli Institute for Particle Astrophysics and Cosmology and Department of Physics, Stanford University, Stanford, CA 94305, USA}
\affiliation{Department of Particle Physics and Astrophysics, SLAC National Accelerator Laboratory, Stanford, CA 94305, USA}

\author{Zhongxu Zhai}
\affiliation{Center for Cosmology and Particle Physics, Department of Physics, New York University, 4 Washington Place, New York, NY 10003, USA}

\begin{abstract}
Existing models for the dependence of the halo mass function on cosmological parameters will become a limiting source of systematic uncertainty for cluster cosmology in the near future. We present a halo mass function emulator and demonstrate improved accuracy relative to state-of-the-art analytic models. In this work, mass is defined using an overdensity criteria of 200 relative to the mean background density.  Our emulator is constructed from the {\sc Aemulus} simulations, a suite of 40 $N$-body simulations with snapshots from $z=3$ to $z=0$. These simulations cover the flat $w$CDM parameter space allowed by recent Cosmic Microwave Background, Baryon Acoustic Oscillation and Type Ia Supernovae results, varying the parameters $w$, $\Omega_m$, $\Omega_b$, $\sigma_8$, $N_{\text{eff}}$, $n_s$, and $H_0$. We validate our emulator using five realizations of seven different cosmologies, for a total of 35 test simulations. These test simulations were not used in constructing the emulator, and were run with fully independent initial conditions. We use our test simulations to characterize the modeling uncertainty of the emulator, and introduce a novel way of marginalizing over the associated systematic uncertainty.  We confirm non-universality in our halo mass function emulator as a function of both cosmological parameters and redshift. Our emulator achieves better than 1\% precision over much of the relevant parameter space, and we demonstrate that the systematic uncertainty in our emulator will remain a negligible source of error for cluster abundance studies through at least the LSST Year 1 data set. 
\end{abstract}


\section{Introduction}
\label{sec:introduction}

The dependence of the halo mass function --- i.e., the comoving density of dark matter halos per unit mass --- on cosmological parameters enables the abundance of galaxy clusters to be an exceptionally promising dark energy probe \citep{detf}. Indeed, the recent \citet{cosmicvisions} report shows that galaxy clusters could provide the tightest constraints yet on cosmological parameters, provided the associated systematic uncertainties can be adequately controlled. Current optical surveys such as the Dark Energy Survey \citep[DES][]{DES05.1,DES15.1,Melchior2017}, the Hyper Suprime-Cam Survey \citep{HSC12,HSCDR1}, and the Kilo-Degree Survey \citep{KiDSDR3,Radovich2017} can expect to observe $\sim$100,000 galaxy clusters, with this number growing by up to order of magnitude in the LSST era \citep{Weinberg13}. SZ and X-ray cluster catalogs \citep[e.g.][]{Bleem2015, Piffaretti2011} will reach similarly large samples of clusters. Here, we take an important step towards calibrating the cosmology dependence of the halo mass function at a precision that is sufficiently high to ensure this source of systematic uncertainty remains negligible in the LSST era.

The largest source of uncertainty in cluster abundance studies is generally cluster mass calibration. Observationally, cluster masses are estimated using a mass--observable relation. The observable can be the thermal SZ effect \citep{Stern2018,Dietrich2017,Bleem2015}, X-ray emission from hot gas \citep{Piffaretti2011,Mehrtens2012}, or optical richness \citep{Melchior2017,Simet2017,Rykoff2016}. The recent analysis of \cite{McClintock2018} in the DES Year 1 data set achieved a 5\% calibration of the mass--richness relation of \redmapper\ clusters. This is expected to improve to $\sim$1\% for LSST and extend out to redshifts $z \approx 1$.

\begin{figure*}[htb!]
	\includegraphics[width=\linewidth]{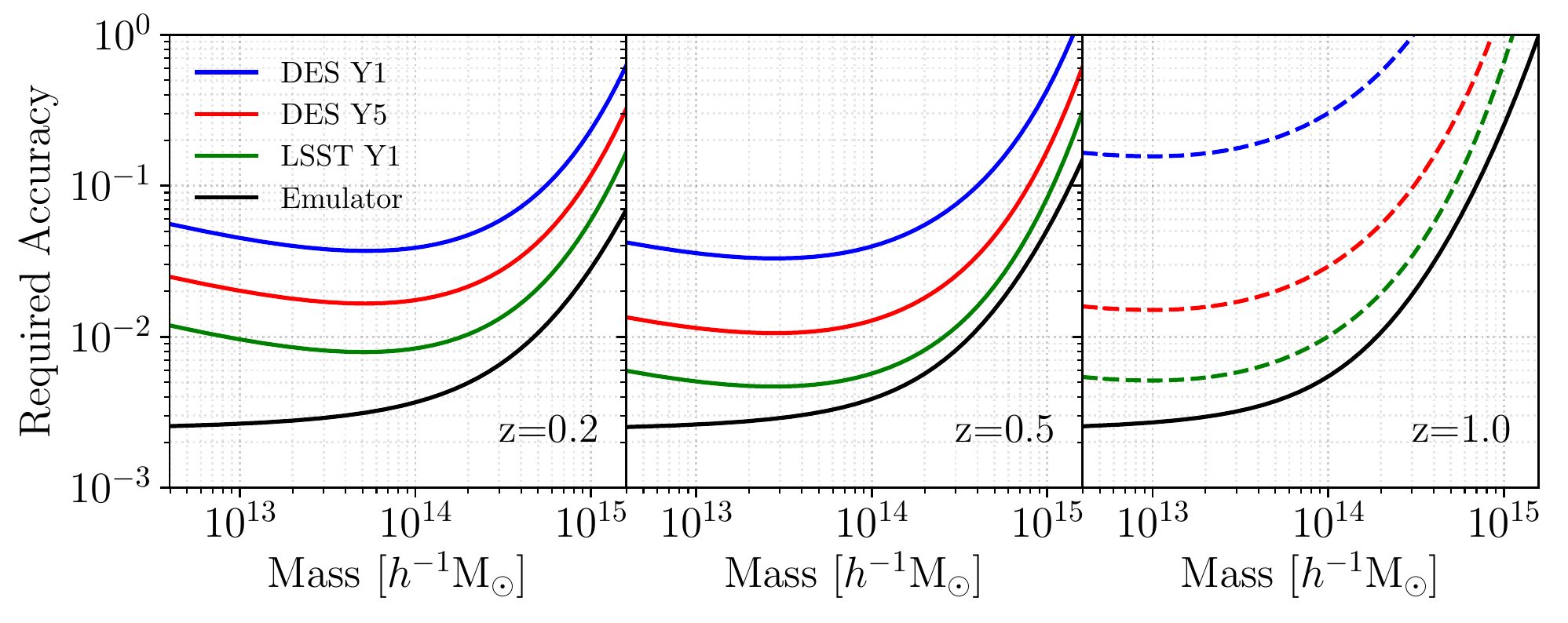}
    \caption{Required accuracy of the halo mass function for different surveys. These accuracies are calculated by demanding that the uncertainty in the halo mass function not increase the error budget of a cluster-abundance measurement by more than 10\% relative to the uncertainty from cluster mass calibration. Mass calibration uncertainties are estimated based on the DES Y1 measurements of DES Y1 cluster ({\it blue}) \citep{McClintock2018}. DES Y5 ({\it red}) and LSST Y1 ({\it green}) requirements are projected based on expected improvements of the weak lensing masses. Dashed lines are extrapolations to $z=1$. The {\it black} curve indicates the current accuracy of the emulator. For details of how the accuracy requirements are computed, see \autoref{app:accuracy_requirements}.}
    \label{fig:accuracy_requirements}
\end{figure*}

In addition to uncertainties in the mass--observable relation, the total error budget in a cluster abundance studies must include a contribution from modeling uncertainty. One such example is the systematic uncertainty associated with modeling the halo mass function. \citet{Reed2007} present a review of halo mass function models published up to that point. A discussion on accuracy requirements for mass function models can be found in \citet{Reed2013}. Assuming clusters map perfectly on to halos, the systematic uncertainty in the calibration of the halo mass function directly translates into a systematic uncertainty in the predicted cluster abundance.  The \citet{Tinker2008} mass function reports an accuracy of $\approx 5\%$ for $10^{11} < M < 10^{15}\ \hmsun$ halos at $z=0$. The calibration was obtained using a suite of flat $\Lambda$CDM simulations that varied the parameters $\Omega_m$, $\Omega_b$, $\sigma_8$, $n_s$, and $H_0$.

\autoref{fig:accuracy_requirements} presents the required accuracy for the systematic uncertainty in the halo mass function to be negligible (quantitatively defined below) in the DES Year 1, DES Year 5, and LSST Year 1 cluster abundance analyses. These requirements are set by rescaling the statistical error reported in \citet{McClintock2018} for the uncertainty in halo mass calibration. Specifically, we  demand that the theoretical uncertainty in the halo mass function must not increase the total error budget of a cluster abundance analysis by more than 10\% relative to the error from mass calibration alone.   The details of this calculation, including assumptions regarding the lens source densities, survey areas, and depths, are provided in \autoref{app:accuracy_requirements}. Importantly, \autoref{fig:accuracy_requirements} shows that the claimed $5\%$ accuracy at $z=0$ of the \citet{Tinker2008} mass function is nearly sufficient for the DES Y1 data set, but not for future data sets. Evidently, a new calibration of the halo mass functions is required to prevent systematic uncertainties in the halo mass function from contributing significantly to the cosmological error budget in cluster abundance studies.

Early halo mass function models were analytic estimates \citep{PressSchechter74}. More recent analyses have calibrated fitting functions from simulations \citep{ShethTormen99,ShethMoTormen01}. \citet{Jenkins01} and \citet{Evrard02} attempted to find a universal fitting function accurate to $\sim 10\%-20\%$, where the mass function did not explicitly depend on cosmological parameters. Later, \citet{Warren06} calibrated a fitting function on a single cosmology at $z=0$ and achieved $\sim 5\%$ accuracy. \citet{Tinker2008} calibrated the halo mass function to $\approx 5\%$ at $z=0$ for virial masses identified using the spherical overdensity (SO) algorithm in the range $10^{11} \leq M \leq 10^{15}\ \hmsun$. \citet{Tinker2008} has been the standard for cluster abundance analyses since then \citep{Zu14,PlanckXXIV2015,Mantz2016}.  Similar high-precision calibrations of the halo mass function for friends-of-friends halos also exist \citep{bhattacharyaetal11}.

More recently, \citet{Heitmann2016} built a high-dimensional interpolator using Gaussian Processes \citep{RasmussenWilliamsGPs} to model the halo mass function for halos identified using a {\it friends-of-friends} (FOF) algorithm. In an 8-dimensional cosmological parameter space they were able to accurately model the mass function for their entire mass range at $z=0$. The choice of halo finding algorithm impacts the recovered halo mass function. In practice, the appropriateness of different halo definitions depends on the specific science question under consideration. In this work, we have chosen to focus exclusively on SO halos. Note that the connection between FOF and SO halos is non-trivial, and exhibits strong asymmetric scatter \citep{More2011,Tinker2008}. Thus we caution that our results are only appropriate for the specific mass definition we have adopted.

The focus of this paper is to present a halo mass function emulator calibrated specifically for SO halos in numerical simulations. For the purposes of this work, halo mass is defined using an over-density threshold $\Delta=200\text{b}$, which is 200 times the mean matter density of the Universe at the epoch at which the halo is identified. Our approach involves emulating parameters in a fitting function rather than $n(M,z)$ directly. This approach has an advantage over directly emulating the mass function in that it requires significantly less training data. Moreover, because our fitting functions are expressed in terms of the peak-height $\nu$, much of the cosmological dependence of the mass function is already removed. This way, the emulator need only characterize any remaining cosmological sensitivity.

The original mass function by \citet{PressSchechter74} was universal when written in terms of the peak height $\nu$. That is, the authors proposed a form of the mass function that did not explicitly depend on cosmological parameters nor did it evolve over time. Recently, \citet{Despali2016} found that the mass function using the $M_{\rm vir}$ definition is universal at the few percent level, and that other definitions such as $M_{\rm 200b}$ are less universal. For constant overdensities $\Delta$ \citet{Tinker2008} demonstrated non-universality at the $20-50\%$ level between different epochs in their simulations. Our emulator incorporates non-universality by design, and we demonstrate the dependence of the emulated halo multiplicity function on redshift, $\Omega_m$ and $\sigma_8$. We note that different definitions of halo mass may lead to increased or decreased universality.   

We quantitatively characterize the performance of our emulator {\it a-posteriori} and provide a model for the residuals of the halo mass function emulator. This residual model accurately describes the non-constant uncertainty over the masses and epochs in our simulations, allowing for the uncertainty in the mass function emulator to be properly propagated into abundance analyses.  We find that our emulator achieves sub-percent accuracy for certain mass ranges, and is accurate enough for the LSST Y1 analysis out to at least $z=1$. 

The layout of this paper is as follows. In \autoref{sec:simulations}, we briefly discuss the simulations used in this work, detail the process of identifying halos, and define the fitting function used. \autoref{sec:emulator_design} describes our method for training the Gaussian Processes and constructing the halo mass function emulator. In \autoref{sec:emulator_accuracy} we show tests of the accuracy of the emulator, while in \autoref{sec:modeling_the_residuals} we present a model for the residuals of the comparison between the emulator predictions and the measured halo mass functions. \autoref{sec:mass_function_universality} contains a discussion of non-universality in the emulator for the halo mass function. \autoref{sec:comparison_to_other_simulations} shows the emulator predictions for additional simulations that probe both lower and higher masses than those used to construct our emulator. \autoref{sec:conclusions} summarizes our results. In \autoref{app:verifying_em_unc_model} we present tests of the residual model, while in \autoref{app:accuracy_requirements} we derive the accuracy requirements for current and near-future cluster cosmology analyses. We make our simulations available at \url{https://AemulusProject.github.io}.


\section{Simulations}
\label{sec:simulations}

Massive clusters live in the high-mass exponentially decaying tail of the halo mass function. Properly investigating this regime requires $N$-body simulations with enough resolution and volume to ensure sufficient statistics for high-mass halos. Collectively, all our simulations comprise the \aemulus\ suite \citep{DeRose2017}. The ``training simulations'' used to construct the emulator consist of 40 $N$-body simulations. Each simulation has a length $L=1050\ \hmpc$ with periodic boundary conditions and $1400^3$ particles. The cosmologies of these simulations span the allowed $4\sigma$ parameter space spanned by a union of {\it Planck} and WMAP9 \citep{PlanckXVI,WMAP9} plus BAO from BOSS \citep{Anderson2014} and the Union 2.1 SNIa data \citep{Suzuki2012}. Initial conditions vary between each simulation, and are specified using \textsc{CAMB}. As described in \cite{DeRose2017}, the cosmological sampling of the simulations was designed using an orthogonal Latin hypercube design \citep{Heitmann2009} and they have an effective volume of 42 $(\hgpc)^3$. Particle masses are given by $M_{\rm part} = 3.513\times 10^{10}(\frac{\Omega_m}{0.3})\ \hmsun$.

In order to assess the performance of the emulator we ran another set of 35 simulations, dubbed the ``test simulations''. These simulations are comprised of seven different cosmologies with five different sets of initial conditions per cosmology. The five realizations are combined in order to reduce sample variance when validating the emulator performance. None of the test simulations were used in the construction of the emulator, and they span the range of cosmological parameter space used to define the training simulations.


\subsection{Cosmological Models}
\label{sec:cosmological_models}

\begin{figure}[htb!]
	\includegraphics[width=\linewidth]{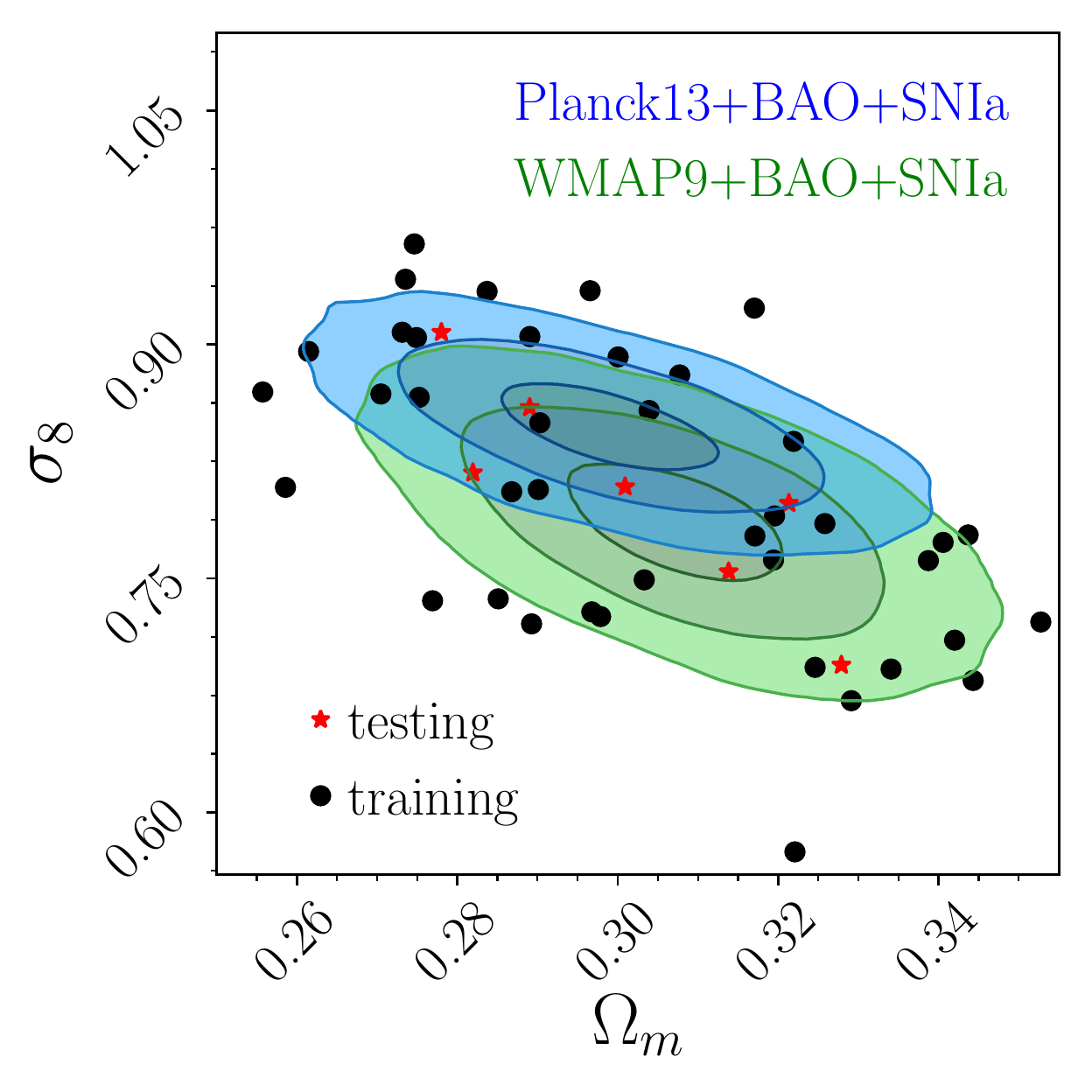}
    \caption{The CMB+BAO+SNIa allowed parameter space (contours) for $\sigma_8$ and $\Omega_M$. This is a combination of BAO from BOSS DR11, the Union 2.1 SNIa catalog, and {\it Planck}/WMAP9. Contour levels are the 1, 2, and 3$\sigma$ confidence contours. {\it Points} are the locations of the training simulations used to construct the emulator.  The red stars mark the locations of the test simulations.}
    \label{fig:simulations}
\end{figure}

Cluster abundance is most sensitive to the matter power spectrum normalization as specified by $\sigma_8$ and to the matter density $\Omega_M$. The training simulations exist in the parameter space $p\in [{\Omega_bh^2, \Omega_ch^2, w, n_s, H_0, N_{\text{eff}},\sigma_8}]$ where $\Omega_b$ is the baryonic matter fraction, $\Omega_c$ is the cold dark matter fraction, $n_s$ is the power spectrum index, $h=H_0(100\ {\rm km\ s^{-1}\ Mpc^{-1}})^{-1}$  is the Hubble constant, and $N_{\text{eff}}$ is the effective number of relativistic species. Both the training and test simulations are shown as points in \autoref{fig:simulations} overlaid on top of the likelihood contours they are designed to span.

\subsection{Halo Identification}
\label{sec:halo_identification}

Halos were identified using the {\sc Rockstar} halo finder \citep{Behroozi2013}, which identifies halos across simulation snapshots. We use the $M_{\rm 200b}$ mass definition, where the halo is defined as a spherical overdensity (SO) $\Delta=200$ times more dense than the background. We conservatively only consider halos with 200 or more particles. The mass and abundance of the halos with the smallest particle numbers (below 500-1000 particles) were found to depend on the mass resolution of the simulations. To account for this systematic, we applied a correction to the recovered abundances as described in Section 4.2.5 of \citealt{DeRose2017}. 

Halos in each snapshot were split into mass bins beginning at the minimum halo mass resolved in each simulation. The maximum edge was fixed arbitrarily at $10^{17}\ \hmsun$, however the maximum halo mass in the simulations was $7.06\times 10^{15}\ \hmsun$. Subhalos were ignored. Using $8^3=512$ spatial jackknife subregions, we estimated the covariance matrix between bins in a given snapshot. We ignore correlations between mass bins across different snapshots when performing the fits described in \autoref{sec:mass_function}.


\subsection{Mass Function}
\label{sec:mass_function}

Our emulators were not trained on the measured mass functions directly. Instead, we fit the mass function of each simulation snapshot with a modified version of the mass function presented in Appendix C of \citet{Tinker2008}. Similar fitting functions were presented in \citet{Jenkins01} and \cite{Warren06}. In \citet{Tinker2008}, the cosmological parameters altered the mass function in two ways: 1) they alter the contribution of matter to the critical density $\Omega_m\rho_c$ and 2) they change the mapping from mass to the RMS variance of the linear density field $\sigma(M,z)$. We extend this approach by allowing the fitting function parameters to have cosmological dependence as well, which is captured using the Gaussian Processes that underpin the emulator.

This fitting function has the following form:
\begin{equation}
	\label{eq:mf_form}
	\frac{{\rm d}n}{{\rm d}M} = G(\sigma)\frac{\bar{\rho}_m}{M}\frac{{\rm d}\ln\sigma^{-1}}{{\rm d}M},
\end{equation}
where the halo multiplicity function $G(\sigma)$ is given by
\begin{equation}
	\label{eq:multiplicity}
	G(\sigma) = B\left[\left(\frac{\sigma}{e}\right)^{-d} + \sigma^{-f} \right]\exp{(-g/\sigma^2)}.
\end{equation}
Here, $\sigma^2$ is the rms variance of the linear density field
\begin{equation}
	\label{eq:sigma_definition}
	\sigma^2 = \int \frac{{\rm d}k}{k}\ \frac{k^3P(k,z)}{2\pi^2}\hat{W}(kR)\,
\end{equation}
evaluated at the Lagrangian scale of the halo, i.e. $R = (3M/4\pi\bar{\rho}_m)^{1/3}$. $P(k,z)$ is the linear matter power spectrum as a function of wavenumber $k$ and redshift $z$, and $\hat{W}$ is the Fourier transform of the real-space top-hat window function. Additionally, we require all dark matter resides in halos, which implies
\begin{equation}
	\label{eq:all_matter_in_halos}
	\int {\rm d}\ln \sigma^{-1} G(\sigma) = 1\,.
\end{equation}

The simulation is unable to sample arbitrarily large modes of the power spectrum due to the finite size of the box. We confirmed that our results are insensitive to a cut in $k=2\pi/R$ at the scale of our simulation $R=1.05\ \hgpc$. We use the publicly available \textsc{CLASS}\footnote{\url{http://class-code.net/}} to calculate the power spectrum. We note that attempting to use analytic transfer functions resulted in a clear decrease in performance by the emulator. By solving \autoref{eq:all_matter_in_halos} we can write $B$ in terms of the remaining free parameters in \autoref{eq:multiplicity}:
\begin{equation}
	\label{eq:B_definition}
    B = 2\left[e^dg^{-d/2}\Gamma\left(\frac{d}{2}\right) + g^{-f/2}\Gamma\left(\frac{f}{2}\right) \right]^{-1}.
\end{equation}
The fitting function parameters are $d$, $e$, $f$, and $g$. The parameter $e$ sets the amplitude and $d$ and $f$ set the slope of low-mass power law; $g$ determines the cutoff mass where the abundance of halos decreases exponentially. Empirically we find that for every cosmology tested, the redshift evolution of the best-fit parameters is well described as a linear function of the scale factor $a$. Therefore, in our analysis we simultaneously fit all snapshots of each simulation simultaneously enforcing linear evolution of the fitting parameters. Specifically, we set:
\begin{equation}
	\label{eq:parameter_scale_factor_dependence}
	p(\Omega,a) = p_0(\Omega) + (a-0.5)p_1(\Omega).,
\end{equation}
In this equation, $p_0$ is value of the fitting function parameter at a scale factor of $a=0.5$ and $p_1$ is the slope of the parameter as a function of the scale factor. We found that fixing $d_0=2.4$ and $f_1=0.12$ optimized the performance of our emulator while removing degeneracies amongst our full parameter set. We still allow $d_1$ and $f_0$ to vary as a function of cosmological parameters. Note that by construction the Gaussian Processes only interpolated over cosmological parameter space and not over redshift and mass.


\subsection{Fitting the training Simulations}
\label{sec:fitting_the_simulations}

After measuring the mass function in each simulation, we found the best-fit model using the fitting function in \autoref{eq:multiplicity}. The best-fit parameters are later interpolated across cosmological parameter space using Gaussian Processes in order to create the emulator. Our best-fit models are obtained by maximizing the likelihood
\begin{equation}
	\label{eq:likelihood}
    \ln\mathcal{L} = -\frac{1}{2}\sum_{i=0}^{9} \left({\bf N}(z_i) - {\bf N}_{\rm model}(z_i)\right)^T {\bf C}_{N,i}^{-1} \left({\bf N}(z_i) - {\bf N}_{\rm model}(z_i)\right)
\end{equation}
where $i$ indexes the ten snapshots at each of the $z_i$ redshifts, ${\bf C}_{N,i}$ is the jackknife-estimated covariance matrix of the abundances at snapshot $i$, and ${\bf N}(z_i)$ is the mass function data at each mass bin in that snapshot. When inverting ${\bf C}_{N,i}$ we applied the correction from \citet{Hartlap07}. The measured quantity is the number of halos in a given ($j$th) mass bin $N_j$. The mass bins are not small, so they are modeled by
\begin{equation}
	\label{eq:mass_function_in_bin}
    N_{{\rm model},j} = V\int_{M_{{\rm min},j}}^{M_{{\rm max},j}}{\rm d}M\ \frac{{\rm d}n}{{\rm d}M},
\end{equation}
where $V=1.05^3\ (\hgpc)^3$ is the volume the simulation, and $M_{{\rm min},j}$ ($M_{{\rm max},j}$) is the minimum (maximum) mass in the $j$th bin. Choosing the number of bins meant striking a balance between obtaining as much resolution as possible while keeping the covariance matrix small enough to estimate by using a spatial jackknife. We used the MCMC code {\it emcee} \citep{Foreman13} to obtain full posterior probability distributions for each of the parameters $d_1$, $e_0$ $e_1$, $f_0$, $g_0$, and $g_1$ for each simulation, imposing flat priors on each of the parameters.


\section{Emulator Design}
\label{sec:emulator_design}

The emulator uses Gaussian Processes to perform regression on the parameters $d_1$, $e_0$, $e_1$, $f_0$, $g_0$ and $g_1$ as a function of the cosmological parameters discussed in \autoref{sec:cosmological_models}. In total the emulator is comprised of six independent Gaussian Processes. Since the Gaussian Processes are independent, we had to rotate the parameters into an orthogonal basis. After doing so each new parameter was labeled $d_1'$, $e_0'$, $e_1'$, $f_0'$, $g_0'$, and $g_1'$. To achieve this we took the MCMC chain from the central-most simulation, index 34, and computed the covariance matrix of the parameters in this chain. Then, by computing its eigenvectors we constructed a rotation matrix that diagonalizes this covariance matrix. We applied the rotation matrix to the MCMC chains for each simulation, so that the parameters of the new chains were all approximately orthogonal. We tested that the rotation matrix used depended only weakly on cosmology, and that the choice of which simulation was used to construct the rotation matrix did not influence the performance of the emulator.

Each Gaussian Process uses the mean and variance of the posterior probability distributions of the orthogonal parameters to interpolate across cosmological parameter space. When the Gaussian Process was used to predict parameters at a new cosmology, the predicted values were rotated back to the original basis.


\subsection{Gaussian Process Kernel}
\label{sec:gaussian_process_kernel}

Consider a set of $n$ measured data $y_i\ i\in[1,...,n]$ with variance $\sigma_i^2$, located at $\vec{x}_i$. If the covariance between two measurements $y_i$ and $y_j$ depends only on $\vec{x}_i$ and $\vec{x}_j$, then the measurements can be approximated by a Gaussian Process $y=\mathcal{GP}(\mu(\vec{x}),{\bf C})$, where $\mu$ is the ``mean function'' and ${\bf C}$ is the covariance matrix. As described in \citet{RasmussenWilliamsGPs}, modeling some data as a Gaussian Process amounts to correctly modeling the covariance, since the mean of the observations $y$ can be subtracted off such that $\mu = 0$. Once a set of data is successfully modeled in this fashion, the Gaussian Process can be used for interpolation. Thus, our approach is to find an optimal covariance matrix describing the mass function parameters as a function of cosmology.  This allows us to interpolate from the locations in parameters space corresponding to our simulations to arbitrary cosmologies. We use the Gaussian Process implementation in the \textsc{Python} package \texttt{george}\footnote{\url{http://george.readthedocs.io/en/latest/}} \citep{Ambikasaran2014}.

We model the covariance matrix as
\begin{equation}
	\label{eq:gp_covariance}
    {\bf C} = {\bf K} + {\bf S}
\end{equation}
where ${\bf K}$ is the kernel matrix and ${\bf S}$ is a diagonal matrix containing the variance of the training data $\sigma^2_{\vec{x}_i}$. Each element of ${\bf K}$ contains the covariance between two data points given by a kernel function. The Gaussian Processes were constructed using a squared-exponential kernel:
\begin{equation}
	\label{eq:emulator_kernel}
	k(\vec{x},\vec{x}') = k_0\exp\left(-\sum_{i}^{N_p}\frac{(x_i-x_i')^2}{2L_i^2}\right)\,,
\end{equation}
where $L_i$ is the length scale of the $i$th cosmological parameter, $k_0$ is the kernel amplitude, and $N_p$ is the number of cosmological parameters. We found fixing $k_0=1$ to be an optimal configuration of the emulators in that residuals of the emulator prediction were minimized. In this approach the $L_i$'s are the hyperparameters of the kernel that we seek to optimize in order to model the covariance of the data. Their values are found by maximizing the likelihood
\begin{equation}
	\label{eq:gp_likelihood}
    \ln\mathcal{L} = -\frac{1}{2}\Delta p^{\rm T}{\bf C}^{-1}\Delta p - \frac{1}{2}\ln \det {\bf C}
\end{equation}
where $\Delta p = p - p_{\rm emu}$. That is, $\Delta p$ is the difference between the measured mass function parameters $p\in \{d_1',e_0',e_1',f_0',g_0',g_1'\}$ and the predictions from the emulator. In attempting to include $k_0$ as a hyperparameter to optimize along with the $L_i$s allowed the Gaussian Processes too much flexibility despite maximizing the likelihood in \autoref{eq:gp_likelihood}. This resulted in poor performance in the tests discussed in \autoref{sec:emulator_accuracy}.


\section{Emulator Accuracy}
\label{sec:emulator_accuracy}

We tested the overall accuracy of our emulator using two sets of tests: leave-one-out tests and predictions of the mass function measured in the test simulations. In both of these tests, the quantity we calculated was the residual difference between the emulator prediction and the observed number of halos in a given mass bin.  Specifically,
\begin{equation}
	\label{eq:residual_definition}
    R = \frac{n - n_{\rm emu}}{n_{\rm emu}}\,.
\end{equation}
In order to compare the emulator performance across simulations of different volumes, we have switched to halo number densities $n$ rather than the actual number of halos $N$.


\subsection{Leave-one-out Tests}
\label{sec:leave_one_out}

\begin{figure*}[htb!]
	\label{fig:all_residuals}
    \includegraphics[width=\linewidth]{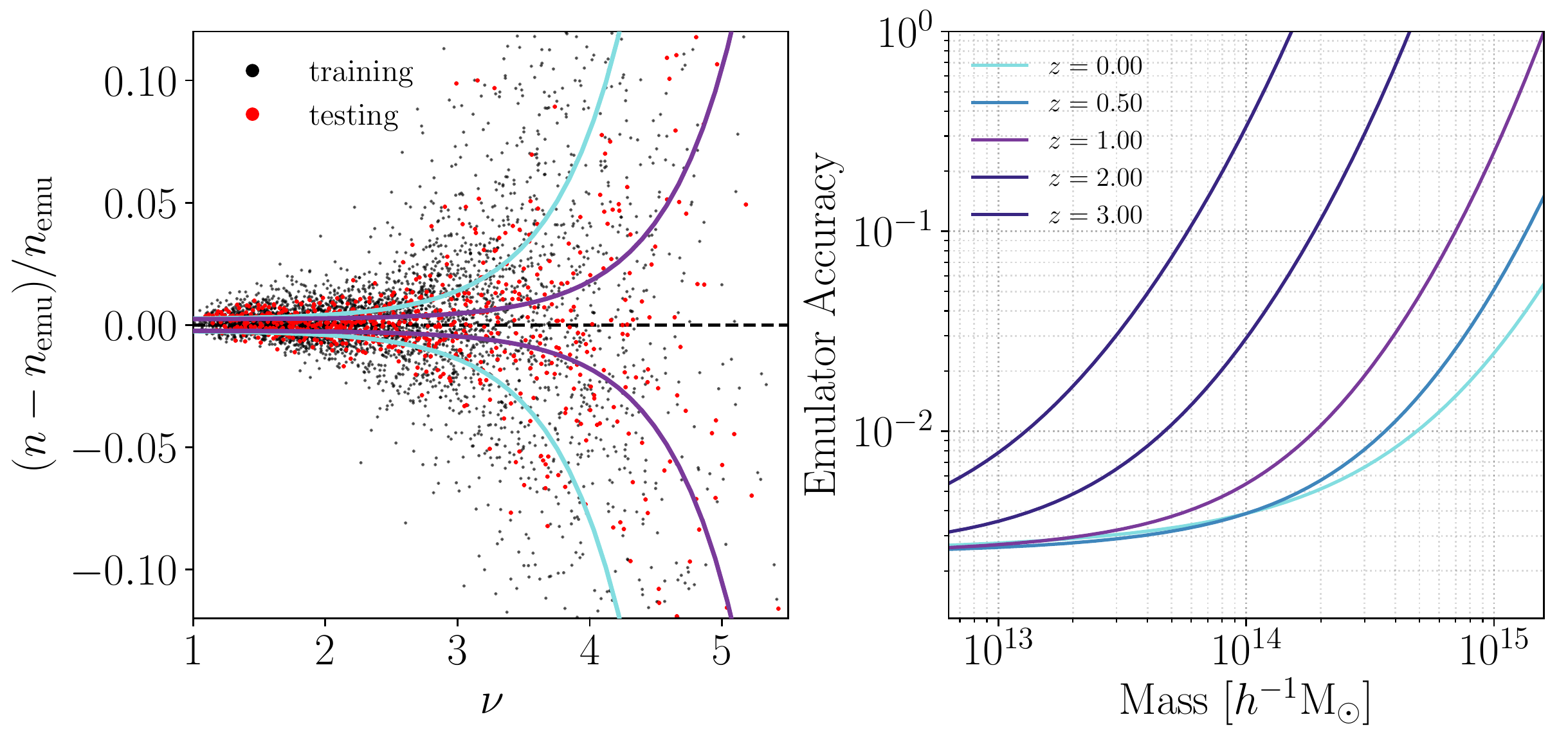}
    \caption{{\it Left}: Residual difference between the emulator prediction and the measured mass function for each mass bins in all snapshots of all of the training and test simulations. The light and dark lines indicate the modeled emulator uncertainty (\autoref{eq:residual_model}) at $z=[0,1]$, respectively. Note at fixed peak height, the emulator precision improves with increasing redshifts. {\it Right}: The modeled emulator accuracy at five redshifts as a function of mass. At fixed mass the emulator accuracy degrades with increasing redshift due to the decreasing halo density with increasing redshift.  By contrast, when working at peak height, the opposite trend is seen: higher precision at increasing redshift.  The difference reflects the redshift-dependent mapping between mass and peak height. For more details about fitting and verifying the model for the emulator accuracy, see \autoref{app:verifying_em_unc_model}.}
\end{figure*}

The leave-one-out tests consist of retraining the Gaussian Processes leaving out the $i$th simulation, and then attempting to predict its mass function. The black points in the left panel of \autoref{fig:all_residuals} show the fractional difference between the measurement and emulator prediction from these tests. The inverse-variance weighted mean of the absolute value of the residuals is $0.57\%$. These tests are Poisson limited past the exponential cutoff at $\nu \approx 3$, where the residuals begin to fan out. The mass of the exponential cutoff depends on the redshift and cosmological parameters.

\subsection{Validation on the Test Simulations}
\label{sec:validation_on_test_sims}

We test the performance of our emulator with the independent test simulations.  We first calculated the average mass function for all simulations that used a single cosmology. Within a given snapshot we calculated the average of the five jackknife estimates of the covariance matrix between the mass bins, and divided this by a factor of five to arrive at the covariance matrix for the mean mass function.

\begin{figure*}[htb!]
	\label{fig:testbox_fit}
    \includegraphics[width=\linewidth]{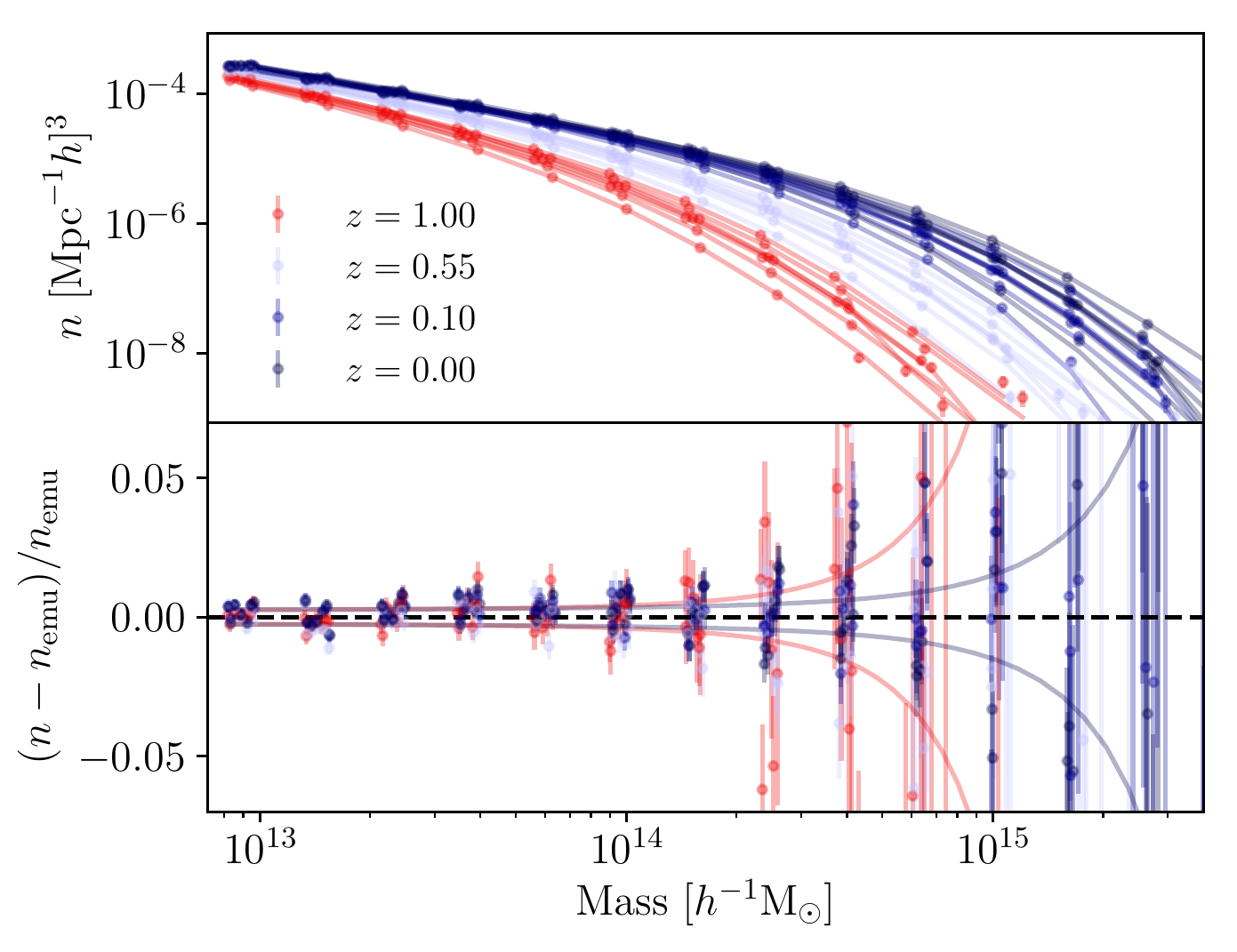}
    \caption{Mass function measurements and emulator predictions for the seven test simulations at four redshifts. These simulations were not used to train the Gaussian Processes in the emulator, and span the entire range of cosmological design space. Points are placed at the mean halo mass of the corresponding bin, and can sometimes scatter left or right when halos are scarce at high masses. Lines in the lower panel show the modeled accuracy predictions from \autoref{eq:residual_model} at $z=[1, 0]$ in red and blue, respectively.}
\end{figure*}

The red points in the left panel of \autoref{fig:all_residuals} show the fractional difference at all redshifts for all test simulations. The inverse-variance weighted mean of the absolute value of the residuals is $0.47\%$, lower than the corresponding value for the leave-one-out tests. \autoref{fig:testbox_fit} shows the prediction and residuals for all the test simulations at four redshifts. The lower panel also shows the modeled emulator uncertainty described below in \autoref{sec:modeling_the_residuals}. 

\begin{figure}[htb!]
	\label{fig:tinker_comparison}
    \includegraphics[width=\linewidth]{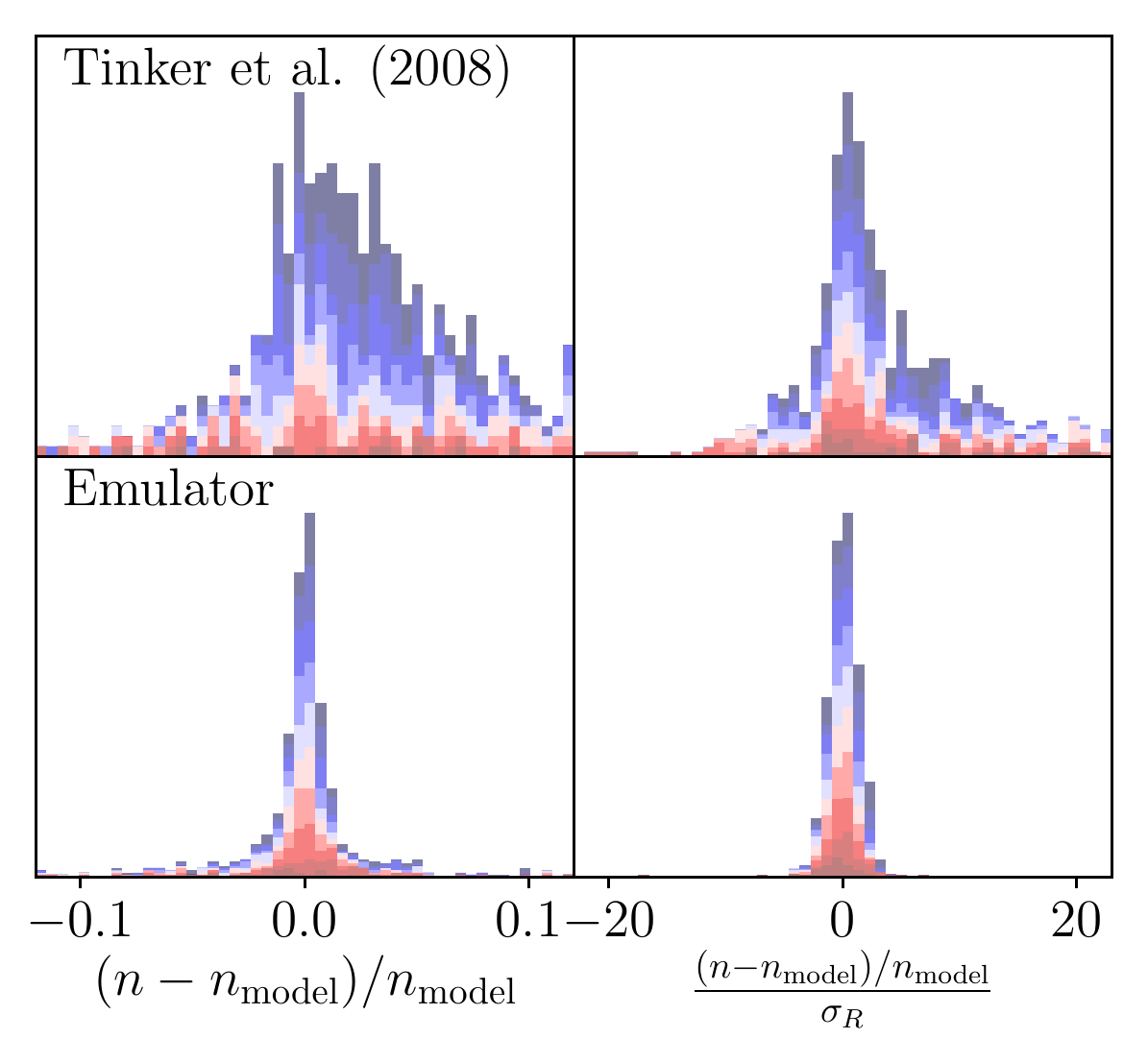} 
    \caption{The residuals from comparing the simulations to the emulator prediction ({\it bottom}) and the \citet{Tinker2008} model ({\it top}) colored by redshift, where red is high redshift and blue is low redshift. The {\it left} panels show the fractional differences, while the {\it right} panel show the distribution of $\chi$. The residual model has not been included in the denominator of the right panels, in comparison to \autoref{eq:residual_chi}.}
\end{figure}

We also test the accuracy of the \citet{Tinker2008} fitting function for the test simulations and compare it to the emulator in \autoref{fig:tinker_comparison}. The top panels show histograms of the residuals between the simulations and the \citet{Tinker2008} model, while the bottom panels show the emulator, with bars colored by redshift: higher redshifts in red and lower redshifts in blue. The inverse-variance weighted mean of the absolute value of the residuals of the \citet{Tinker2008} model is $3.4\%$, substantially larger than the $0.47\%$ found for the emulator. The \citet{Tinker2008} model performs worse than the emulator at all redshifts, though the accuracy is consistent with $\sim 5\%$ at $z=0$. At high redshifts, their model deviates from the simulations significantly, with differences in excess of 100\% for snapshots at $z=[2,3]$. In contrast, the emulator is much more consistent with the simulation, achieving sub-percent accuracy for all snapshots. Though not apparent from the \autoref{fig:tinker_comparison}, the accuracy of the \citet{Tinker2008} model depends on the cosmological parameters of the simulation, unlike the emulator. This is not unexpected, since the \citet{Tinker2008} model was calibrated using simulations in a lower-dimensional parameter space. We performed an identical analysis on the mass function to that presented in \citet{Tinker10} and found comparable results.

\section{Modeling the Residuals}
\label{sec:modeling_the_residuals}

The leave-one-out tests and comparisons to the test simulations demonstrate that the emulator accurately models the mass function. Additionally, \autoref{fig:all_residuals} shows that the level of accuracy is mass dependent. We model this mass dependence using the residuals from the test simulations as defined in \autoref{eq:residual_definition}. The covariance between the residuals of a single mass bin in the $i$th snapshot in a given simulation is
\begin{equation}
	\label{eq:residual_covariance}
    {\bf C}_{R,i} = \frac{{\bf C}_{n, i}}{n_{\rm emu}^2}\,,
\end{equation}
where ${\bf C}_{n} = {\bf C}_{N}/V^2$ is the covariance between the cluster counts used in \autoref{eq:likelihood} divided by the square of the volume. As before, we ignore covariance between snapshots from the same simulations.

The left panel of \autoref{fig:all_residuals} demonstrates how the residuals depend on peak height and redshift. We model the mass-dependent emulator uncertainty, or residual model, using a power law function in $\nu$ and scale factor $a = 1/(1+z)$ given by
\begin{equation}
	\label{eq:residual_model}
    \sigma_{\rm model}(\nu, z) = A_R + 10^{b_R + c_R\widetilde{\nu} + d_R\widetilde{z}}\,,
\end{equation}
where $\widetilde{\nu} = \nu - 3$ and $\widetilde{z} = \frac{1}{1+z} - 0.5$. The free parameters are $A_R$, $b_R$, $c_R$ and $d_R$. This residual model was added to the diagonal of the covariance matrix in \autoref{eq:residual_covariance}
\begin{equation}
	\label{eq:residual_full_covariance}
    \widetilde{\bf C}_{R, i} = {\bf C}_{R,i} + {\bf I}\sigma_{\rm model}^2\,.
\end{equation}
Additionally, we added in quadrature additional uncertainty along the diagonal equal to half the correction to the abundances described in \autoref{sec:halo_identification}. This only negligibly affected the recovered residual model. The free parameters of the residual model were found by maximizing the likelihood
\begin{equation}
	\label{eq:residual_likelihood}
    \ln\mathcal{L}_{R} = -\frac{1}{2}\sum_i \left( {\bf R}_i^{\rm T}\widetilde{\bf C}_{R,i}^{-1}{\bf R}_i + \ln\det\widetilde{\bf C}_{R,i}\right)\,,
\end{equation}
where ${\bf R}_i$ contains the residuals of the $i$th snapshot of a simulation and the sum runs over all snapshot in all simulations. In this way, we model the residuals of individual snapshots as a Gaussian with zero mean and covariance given by \autoref{eq:residual_full_covariance}.

The residual model evaluated at five redshifts appears in the right panel of \autoref{fig:all_residuals}. At fixed mass the accuracy gets worse with redshift, due to the scarcity of massive halos at high redshift. Notably, the trend of the residual model at fixed peak height as a function of redshift is opposite to the trend for mass. At fixed mass, the accuracy degrades at higher redshifts due to there being fewer halos. The halos that found at high redshift reside in very high peaks of the density field, meaning the model is accurate at high peak height at high redshift. This is due to the nonlinear dependence of $\nu(M)$. 

In order to assess the performance of the residual model, we computed the distribution of
\begin{equation}
	\label{eq:residual_chi}
    \chi_{R} = \frac{(n-n_{\rm emu})/n_{\rm emu}}{\sqrt{\sigma_{R}^2 + \sigma_{\rm model}^2}}\, ,
\end{equation}
and performed numerous tests detailed in \autoref{app:verifying_em_unc_model} to confirm its validity. These included comparing the distribution of $\chi_R$ for different cuts in mass and redshift, as well as computing the $\chi^2$ of individual snapshots. The residual model passed all tests, and we found that six out of 668 mass bins with more than 20 halos were more than $3\sigma$ outliers. Notably, four of the outlier mass bins occurred in the same cosmology.

Modeling the residuals allows us to make random realizations of our model uncertainty. Doing so allows one to propagate the mass-dependent modeling uncertainty forward in an abundance analysis. This is accomplished by drawing random residuals from the recovered model uncertainty. Critically, we do not expect this model uncertainty to oscillate wildly as we vary the peak height $\nu$. Rather, we expect the modeling uncertainties to correspond to large-scale fluctuations, meaning fluctuations in neighboring peak height values and/or neighboring redshifts must be strongly correlated. Given two peak heights $\nu_1$ and $\nu_2$ at two different redshifts $z_1$ and $z_2$, we model the correlation coefficient of the model uncertainties with a simple exponential. That is, we set the covariance matrix of the model uncertainties to be:
\begin{equation}
	\label{eq:residual_realization_correlation}
    C(\nu_1,z_1,\nu_2,z_2) = e^{-\Delta\nu^2-\Delta z^2}\sigma_{\rm model}(\nu_1,z_1)\sigma_{\rm model}(\nu_2,z_2)\,,
\end{equation}
where $\Delta\nu = \nu_1-\nu_2$ and $\Delta z = z_1-z_2$ and we have assumed a correlation length of 1 for $\nu$ and $z$.

With the covariance matrix of \autoref{eq:residual_realization_correlation}, we can model the residual systematics as a Gaussian random field. By defining a grid in $\nu$ and $z$, we can compute the corresponding covariance matrix, and make random realizations along this grid that can be interpolated over in order to arrive at smooth residual functions that fall within the statistically allowed range based on our simulation tests. By performing these random draws as part of an MCMC algorithm, one can effectively marginalize over the theoretical uncertainty in the mass function from our emulator. Our implementation of the emulator includes this capability so that analyses can easily marginalize over the emulator uncertainty.


\section{Mass Function Universality}
\label{sec:mass_function_universality}

The halo mass function is universal if the halo multiplicity function $G(\sigma)$ in \autoref{eq:multiplicity} does not depend explicitly on cosmological parameters or redshift \citep{Jenkins01,Evrard02}. 
\citet{Tinker2008} found the mass function was not universal as a function of redshift, or correspondingly large changes in cosmological parameters. This was true whether they used $M_{\rm 200b}$ halos or those defined by $\Delta_{\rm vir}$. Recently, \citet{Despali2016} presented a fitting function to convert spherical overdensity definitions to  $\Delta_{\rm vir}$, which they found to be a more universal definition than others, at least at $z=0$. However, testing the universality of the mass function is limited by the precision of the simulations, and also may be influenced by baryonic effects \citep[see][]{Bocquet2016}. In practice, whether the mass function is universal or not is not relevant for a cluster abundance analysis as long as a model exists that predicts the mass function adequately well. Our emulator explicitly contains non-universality by allowing the parameters in \autoref{eq:multiplicity} to vary with cosmology.

\begin{figure*}[htb!]
	\label{fig:multiplicity_dependence}
    \includegraphics[width=\linewidth]{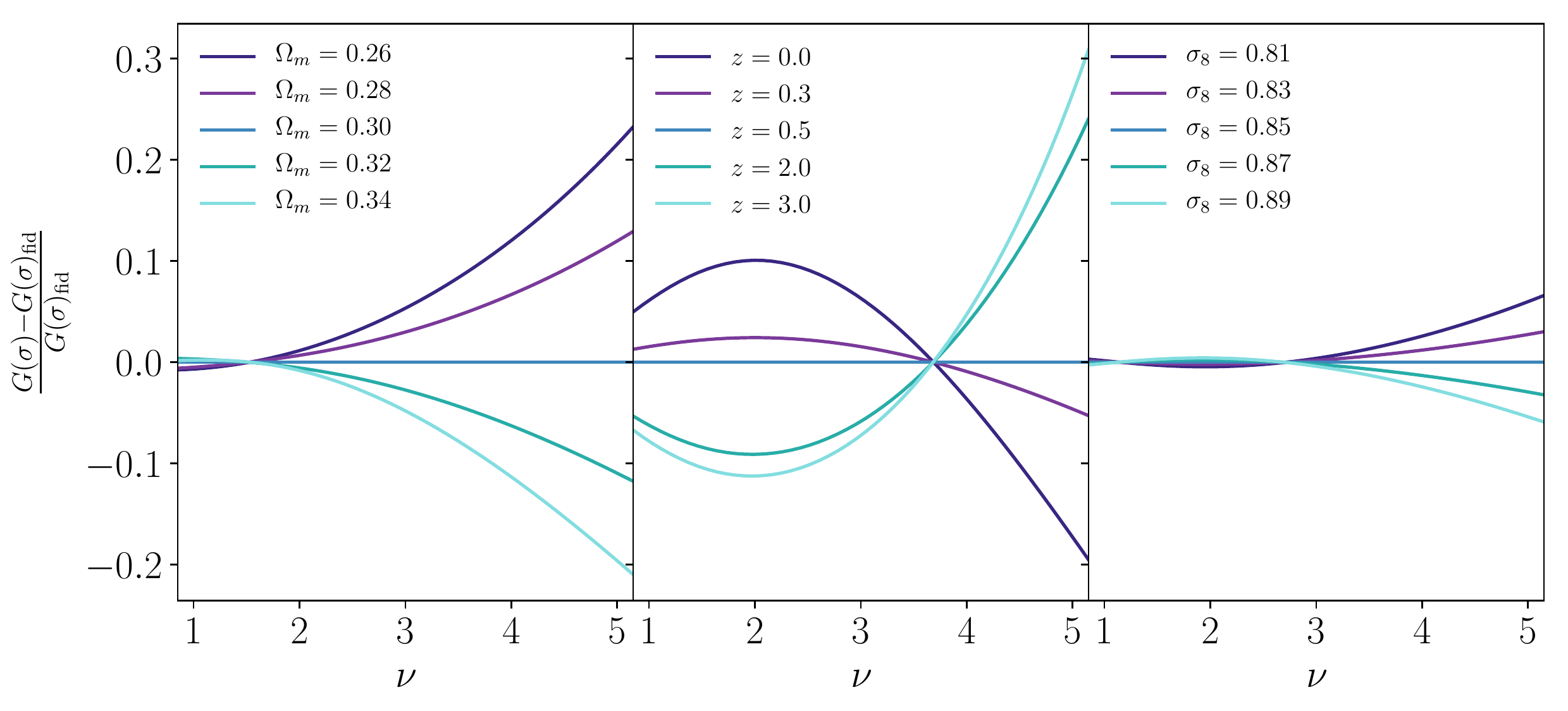}
    \caption{Fractional difference of the halo multiplicity function $G(\sigma)$ as a function of cosmological parameters and redshift compared to a fiducial cosmology at $z=0.5$. If $G(\sigma)$ were universal it would not depend on cosmological parameters or redshift. For clarity the plots are shown as a function of peak height $\nu=\delta_c/\sigma$ where $\delta_c=1.686$.}
\end{figure*}

\autoref{fig:multiplicity_dependence} shows the variation of the multiplicity function in the emulator as a function of $\Omega_m$, $z$ and $\sigma_8$. $G(\sigma)$ depends sensitively on each of these as well as the other cosmological parameters varied in our simulations. The trend of $G(\sigma)$ with $\Omega_m$ are the most straightforward. Recall that, in this panel, all models are constrained to have the same $\sigma_8$ at $z=0$. Thus, at a given peak height---relative to the mean density---structure collapses earlier when the matter density is lower. Earlier collapse leads to higher halo concentration, and a higher mass within $R_{\rm vir}$ at $z=0$. Thus the abundance of halos at fixed $\nu$ increases with lower $\Omega_m$. Similar logical can mostly explain the other trends---halos at fixed $M$ have higher concentrations at lower redshifts, thus higher $G(\sigma)$. And higher values of $\sigma_8$ leads to more early-forming, highly concentrated structure as well. This demonstrates that the halo multiplicity function in our emulator is not universal. This figure serves the same purpose as the sensitivity plots presented in \citet{Heitmann2009}, where they demonstrated the dependence of their power spectrum emulator on cosmological parameters.

\section{Comparison to other Simulations}
\label{sec:comparison_to_other_simulations}

To assess the performance at lower halo masses, we compared the emulator to a set of simulations run with 2048$^3$ particles in volumes 400 $\hmpc$ per side resolving halos down to $\sim 10^{11}\ \hmsun$ with about 200 particles. These are two of our {\it high resolution} suite of simulations. The full high resolution suite is still in progress; it will be comprised of 25 simulations run within the currently allowed parameter space shown in \autoref{fig:simulations}. These are not incorporated into the training data of any of our emulators at present. 

We also ran additional simulations with 3 $\hgpc$ per side and 2048$^3$ particles in order to resolve high-mass halos of mass up to $\sim 5 \times 10^{15}\ \hmsun$. The cosmological parameters of the large boxes were identical to our test simulations to check for consistency across the simulations. We note that we have seen clear evidence that the finite particle correction described in Section 4.2.5 of \citet{DeRose2017} is resolution dependent, and therefore this correction should not be applied to our large volume simulations.  For this reason, when comparing our large volume simulations to our emulator we restrict ourselves to halos with masses $M\geq 3\times 10 10^{14}\ \hmsun$, corresponding to halos with $\geq 1000$ particles.

\begin{figure*}[htb!]
\centering \includegraphics[width=\linewidth]{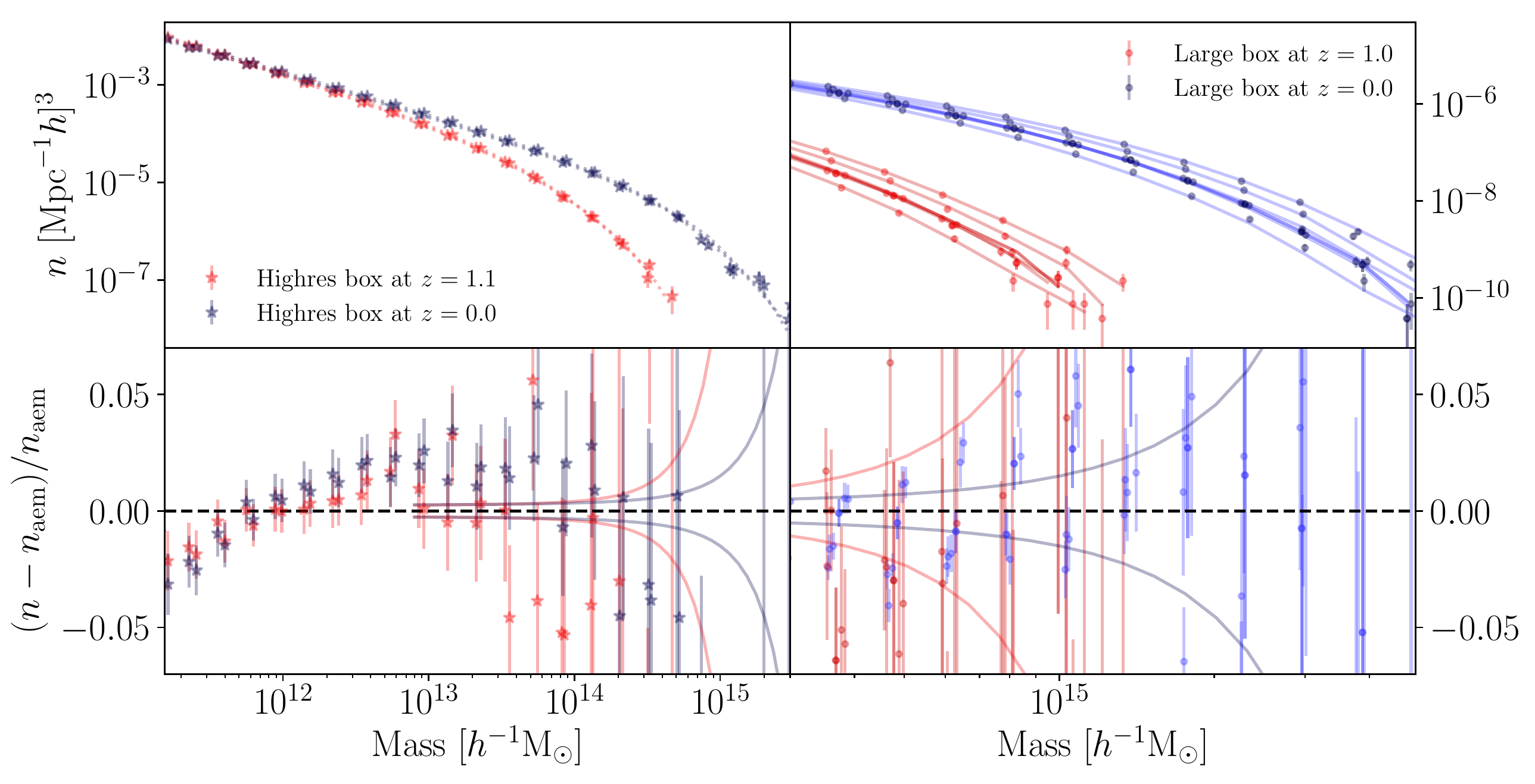}
    \caption{Comparison between the mass function predicted by the emulator and: A) two high resolution boxes with 400 $\hmpc$ on a side ({\it left}), and B) even simulations with 3 $\hgpc$ on a side ({\it right}). Note that the emulator predictions ({\it dotted lines}) for the two high resolution simulations nearly overlap. Lines in the lower panels are the emulator modeling uncertainty from \autoref{eq:residual_model} plotted in the mass range available to the simulations used to train the emulator. The emulator extrapolates well down to low masses.  We conservatively claim an accuracy of 5\% to low masses. At high masses the observed abundances are consistent with the emulator predictions.
    }
    \label{fig:other_simulations}
\end{figure*}

The comparison to these simulations is shown in \autoref{fig:other_simulations}. Our modeling uncertainty (\autoref{sec:modeling_the_residuals}) is only valid for the range of masses used to construct the emulator and may not extrapolate well. However, as can be seen from \autoref{fig:other_simulations}, our emulator extrapolates well down to $\approx 10^{11}\ \hmsun$ at the $\sim 2\%$ level. Due to the possibility of systematics from unconstrained cosmological dependencies and sample variance, we conservatively estimate a precision at these low masses at $\approx 5\%$. At high mass the emulator is consistent with the simulations.

\section{Conclusions}
\label{sec:conclusions}

We present an emulator for the halo mass function constructed from halos identified with spherical overdensities 200 times more dense than the background, the $M_{\rm 200b}$ mass definition. This emulator relies on Gaussian Processes trained on parameters of a fitting function for the halo multiplicity. In this way, non-universality is directly incorporated into the design of the emulator.

We construct the emulator from a suite of 40 simulations. An additional independent set of 35 simulations is used for validation. These test simulations are comprised of five independent simulations of seven different cosmologies, enabling us to reduce the sample variance for each of the seven test cosmologies in the test simulations by a factor of five.

The performance of the emulator is measured with two sets of tests: leave-one-out tests and comparisons to the test simulations. In both tests the emulator successfully predicted the mass function to better than 1\% accuracy in the low mass (i.e. power-law) regime for the redshifts considered. Validation of the accuracy at the high mass end (i.e. the exponential tail) is limited by Poisson noise in the number of halos.  We successfully modeled the overall uncertainty of the emulator across our entire mass range. Finally, we demonstrated in \autoref{fig:accuracy_requirements} that the precision achieved by our emulator is sufficiently high for systematic uncertainties in the halo mass function from dark matter simulations to remain negligible for the DES Year 5 and LSST Year 1 cluster abundance analyses.

\acknowledgments
TM and ER are supported by DOE grant DE-SC0015975. ER acknowledges additional support by the Sloan Foundation, grant FG-2016-6443. JLT and RHW acknowledge support of NSF grant AST-1211889.  JD, RHW, SM, MRB received support from the U.S. Department of Energy under contract number DE-AC02-76SF00515. YYM is supported by the Samuel P.\ Langley PITT PACC Postdoctoral Fellowship. This research used resources of the National Energy Research Scientific Computing Center, a DOE Office of Science User Facility supported by the Office of Science of the U.S. Department of Energy under Contract No. DE-AC02-05CH11231.

\software{Python,
Matplotlib \citep{matplotlib},
NumPy \citep{numpy},
SciPy \citep{scipy},
emcee \citep{Foreman13},
george \citep{Ambikasaran2014},
CAMB \citep{CAMB},
CLASS \citep{CLASS1},
\textsc{Rockstar} \citep{Behroozi2013},
GADGET \citep{Springel05},
2LPT \citep{Crocce2006}
}

\bibliographystyle{yahapj}
\bibliography{astroref,software}

\begin{thebibliography}{}
\providecommand\natexlab[1]{#1}
\providecommand\JournalTitle[1]{#1}

\bibitem[{{Aihara} {et~al.}(2017){Aihara}, {Armstrong}, {Bickerton}, {Bosch},
  {Coupon}, {Furusawa}, {Hayashi}, {Ikeda}, {Kamata}, {Karoji}, {Kawanomoto},
  {Koike}, {Komiyama}, {Lupton}, {Mineo}, {Miyatake}, {Miyazaki}, {Morokuma},
  {Obuchi}, {Oishi}, {Okura}, {Price}, {Takata}, {Tanaka}, {Tanaka}, {Tanaka},
  {Uchida}, {Uraguchi}, {Utsumi}, {Wang}, {Yamada}, {Yamanoi}, {Yasuda},
  {Arimoto}, {Chiba}, {Finet}, {Fujimori}, {Fujimoto}, {Furusawa}, {Goto},
  {Goulding}, {Gunn}, {Harikane}, {Hattori}, {Hayashi}, {Helminiak}, {Higuchi},
  {Hikage}, {Ho}, {Hsieh}, {Huang}, {Huang}, {Imanishi}, {Iwata}, {Jaelani},
  {Jian}, {Kashikawa}, {Katayama}, {Kojima}, {Konno}, {Koshida}, {Kusakabe},
  {Leauthaud}, {Lee}, {Lin}, {Lin}, {Mandelbaum}, {Matsuoka}, {Medezinski},
  {Miyama}, {Momose}, {More}, {More}, {Mukae}, {Murata}, {Murayama}, {Nagao},
  {Nakata}, {Niikura}, {Nishizawa}, {Oguri}, {Okabe}, {Ono}, {Onodera},
  {Onoue}, {Ouchi}, {Pyo}, {Shibuya}, {Shimasaku}, {Simet}, {Speagle},
  {Spergel}, {Strauss}, {Sugahara}, {Sugiyama}, {Suto}, {Suzuki}, {Tait},
  {Takada}, {Terai}, {Toba}, {Turner}, {Uchiyama}, {Umetsu}, {Urata}, {Usuda},
  {Yeh}, \& {Yuma}}]{HSCDR1}
{Aihara}, H., {Armstrong}, R., {Bickerton}, S., {et~al.} 2017,
  \JournalTitle{ArXiv e-prints},
  \href{http://arxiv.org/abs/1702.08449}{{\sffamily arXiv:1702.08449
  [astro-ph.IM]}}

\bibitem[{{Albrecht} {et~al.}(2006){Albrecht}, {Bernstein}, {Cahn}, {Freedman},
  {Hewitt}, {Hu}, {Huth}, {Kamionkowski}, {Kolb}, {Knox}, {Mather}, {Staggs},
  \& {Suntzeff}}]{detf}
{Albrecht}, A., {Bernstein}, G., {Cahn}, R., {et~al.} 2006, \JournalTitle{ArXiv
  Astrophysics e-prints},
  \href{http://arxiv.org/abs/astro-ph/0609591}{{\sffamily astro-ph/0609591}}

\bibitem[{{Ambikasaran} {et~al.}(2015){Ambikasaran}, {Foreman-Mackey},
  {Greengard}, {Hogg}, \& {O'Neil}}]{Ambikasaran2014}
{Ambikasaran}, S., {Foreman-Mackey}, D., {Greengard}, L., {Hogg}, D.~W., \&
  {O'Neil}, M. 2015,
  \href{http://dx.doi.org/10.1109/TPAMI.2015.2448083}{\JournalTitle{IEEE
  Transactions on Pattern Analysis and Machine Intelligence}, 38},
  \href{http://arxiv.org/abs/1403.6015}{{\sffamily arXiv:1403.6015 [math.NA]}}

\bibitem[{{Anderson} {et~al.}(2014){Anderson}, {Aubourg}, {Bailey}, {Beutler},
  {Bhardwaj}, {Blanton}, {Bolton}, {Brinkmann}, {Brownstein}, {Burden},
  {Chuang}, {Cuesta}, {Dawson}, {Eisenstein}, {Escoffier}, {Gunn}, {Guo}, {Ho},
  {Honscheid}, {Howlett}, {Kirkby}, {Lupton}, {Manera}, {Maraston}, {McBride},
  {Mena}, {Montesano}, {Nichol}, {Nuza}, {Olmstead}, {Padmanabhan},
  {Palanque-Delabrouille}, {Parejko}, {Percival}, {Petitjean}, {Prada},
  {Price-Whelan}, {Reid}, {Roe}, {Ross}, {Ross}, {Sabiu}, {Saito}, {Samushia},
  {S{\'a}nchez}, {Schlegel}, {Schneider}, {Scoccola}, {Seo}, {Skibba},
  {Strauss}, {Swanson}, {Thomas}, {Tinker}, {Tojeiro}, {Maga{\~n}a}, {Verde},
  {Wake}, {Weaver}, {Weinberg}, {White}, {Xu}, {Y{\`e}che}, {Zehavi}, \&
  {Zhao}}]{Anderson2014}
{Anderson}, L., {Aubourg}, {\'E}., {Bailey}, S., {et~al.} 2014,
  \href{http://dx.doi.org/10.1093/mnras/stu523}{\JournalTitle{\mnras}, 441, 24}

\bibitem[{{Behroozi} {et~al.}(2013){Behroozi}, {Wechsler}, \&
  {Wu}}]{Behroozi2013}
{Behroozi}, P.~S., {Wechsler}, R.~H., \& {Wu}, H.-Y. 2013,
  \href{http://dx.doi.org/10.1088/0004-637X/762/2/109}{\JournalTitle{\apj},
  762, 109}

\bibitem[{{Bhattacharya} {et~al.}(2011){Bhattacharya}, {Heitmann}, {White},
  {Luki{\'c}}, {Wagner}, \& {Habib}}]{bhattacharyaetal11}
{Bhattacharya}, S., {Heitmann}, K., {White}, M., {et~al.} 2011,
  \href{http://dx.doi.org/10.1088/0004-637X/732/2/122}{\JournalTitle{\apj},
  732, 122}

\bibitem[{{Bleem} {et~al.}(2015){Bleem}, {Stalder}, {de Haan}, {Aird}, {Allen},
  {Applegate}, {Ashby}, {Bautz}, {Bayliss}, {Benson}, {Bocquet}, {Brodwin},
  {Carlstrom}, {Chang}, {Chiu}, {Cho}, {Clocchiatti}, {Crawford}, {Crites},
  {Desai}, {Dietrich}, {Dobbs}, {Foley}, {Forman}, {George}, {Gladders},
  {Gonzalez}, {Halverson}, {Hennig}, {Hoekstra}, {Holder}, {Holzapfel},
  {Hrubes}, {Jones}, {Keisler}, {Knox}, {Lee}, {Leitch}, {Liu}, {Lueker},
  {Luong-Van}, {Mantz}, {Marrone}, {McDonald}, {McMahon}, {Meyer}, {Mocanu},
  {Mohr}, {Murray}, {Padin}, {Pryke}, {Reichardt}, {Rest}, {Ruel}, {Ruhl},
  {Saliwanchik}, {Saro}, {Sayre}, {Schaffer}, {Schrabback}, {Shirokoff},
  {Song}, {Spieler}, {Stanford}, {Staniszewski}, {Stark}, {Story}, {Stubbs},
  {Vanderlinde}, {Vieira}, {Vikhlinin}, {Williamson}, {Zahn}, \&
  {Zenteno}}]{Bleem2015}
{Bleem}, L.~E., {Stalder}, B., {de Haan}, T., {et~al.} 2015,
  \href{http://dx.doi.org/10.1088/0067-0049/216/2/27}{\JournalTitle{\apjs},
  216, 27}

\bibitem[{{Bocquet} {et~al.}(2016){Bocquet}, {Saro}, {Dolag}, \&
  {Mohr}}]{Bocquet2016}
{Bocquet}, S., {Saro}, A., {Dolag}, K., \& {Mohr}, J.~J. 2016,
  \href{http://dx.doi.org/10.1093/mnras/stv2657}{\JournalTitle{\mnras}, 456,
  2361}

\bibitem[{{Crocce} {et~al.}(2006){Crocce}, {Pueblas}, \&
  {Scoccimarro}}]{Crocce2006}
{Crocce}, M., {Pueblas}, S., \& {Scoccimarro}, R. 2006,
  \href{http://dx.doi.org/10.1111/j.1365-2966.2006.11040.x}{\JournalTitle{\mnras},
  373, 369}

\bibitem[{{Dark Energy Survey Collaboration}(2005)}]{DES05.1}
{Dark Energy Survey Collaboration}. 2005, \JournalTitle{ArXiv e-prints},
  \href{http://arxiv.org/abs/astro-ph/0510346}{{\sffamily
  arXiv:astro-ph/0510346}}

\bibitem[{{Dark Energy Survey Collaboration}(2016)}]{DES15.1}
---. 2016, \href{http://dx.doi.org/10.1093/mnras/stw641}{\JournalTitle{\mnras},
  460, 1270}

\bibitem[{{de Jong} {et~al.}(2017){de Jong}, {Kleijn}, {Erben}, {Hildebrandt},
  {Kuijken}, {Sikkema}, {Brescia}, {Bilicki}, {Napolitano}, {Amaro}, {Begeman},
  {Boxhoorn}, {Buddelmeijer}, {Cavuoti}, {Getman}, {Grado}, {Helmich}, {Huang},
  {Irisarri}, {La Barbera}, {Longo}, {McFarland}, {Nakajima}, {Paolillo},
  {Puddu}, {Radovich}, {Rifatto}, {Tortora}, {Valentijn}, {Vellucci}, {Vriend},
  {Amon}, {Blake}, {Choi}, {Conti}, {Gwyn}, {Herbonnet}, {Heymans}, {Hoekstra},
  {Klaes}, {Merten}, {Miller}, {Schneider}, \& {Viola}}]{KiDSDR3}
{de Jong}, J.~T.~A., {Kleijn}, G.~A.~V., {Erben}, T., {et~al.} 2017,
  \href{http://dx.doi.org/10.1051/0004-6361/201730747}{\JournalTitle{\aap},
  604, A134}

\bibitem[{{DeRose}(2018)}]{DeRose2017}
{DeRose}, J. e.~a. 2018, \JournalTitle{submitted to ApJ}

\bibitem[{{Despali} {et~al.}(2016){Despali}, {Giocoli}, {Angulo}, {Tormen},
  {Sheth}, {Baso}, \& {Moscardini}}]{Despali2016}
{Despali}, G., {Giocoli}, C., {Angulo}, R.~E., {et~al.} 2016,
  \href{http://dx.doi.org/10.1093/mnras/stv2842}{\JournalTitle{\mnras}, 456,
  2486}

\bibitem[{{Dietrich} {et~al.}(2017){Dietrich}, {Bocquet}, {Schrabback},
  {Hoekstra}, {Grandis}, {Mohr}, {Allen}, {Bayliss}, {Benson}, {Bleem},
  {Brodwin}, {Bulbul}, {Capasso}, {Chiu}, {Crawford}, {Gonzalez}, {de Haan},
  {Klein}, {von der Linden}, {Mantz}, {Marrone}, {McDonald}, {Raghunathan},
  {Rapetti}, {Reichardt}, {Saro}, {Stalder}, {Stark}, {Stern}, \&
  {Stubbs}}]{Dietrich2017}
{Dietrich}, J.~P., {Bocquet}, S., {Schrabback}, T., {et~al.} 2017,
  \JournalTitle{ArXiv e-prints},
  \href{http://arxiv.org/abs/1711.05344}{{\sffamily arXiv:1711.05344}}

\bibitem[{{Dodelson} {et~al.}(2016){Dodelson}, {Heitmann}, {Hirata},
  {Honscheid}, {Roodman}, {Seljak}, {Slosar}, \& {Trodden}}]{cosmicvisions}
{Dodelson}, S., {Heitmann}, K., {Hirata}, C., {et~al.} 2016,
  \JournalTitle{ArXiv e-prints},
  \href{http://arxiv.org/abs/1604.07626}{{\sffamily arXiv:1604.07626}}

\bibitem[{{Evrard} {et~al.}(2002)}]{Evrard02}
{Evrard}, A.~E., {et~al.} 2002, \JournalTitle{\apj}, 573, 7

\bibitem[{{Foreman-Mackey} {et~al.}(2013){Foreman-Mackey}, {Hogg}, {Lang}, \&
  {Goodman}}]{Foreman13}
{Foreman-Mackey}, D., {Hogg}, D.~W., {Lang}, D., \& {Goodman}, J. 2013,
  \href{http://dx.doi.org/10.1086/670067}{\JournalTitle{\pasp}, 125, 306}

\bibitem[{{Hartlap} {et~al.}(2007){Hartlap}, {Simon}, \&
  {Schneider}}]{Hartlap07}
{Hartlap}, J., {Simon}, P., \& {Schneider}, P. 2007,
  \href{http://dx.doi.org/10.1051/0004-6361:20066170}{\JournalTitle{\aap}, 464,
  399}

\bibitem[{{Heitmann} {et~al.}(2009){Heitmann}, {Higdon}, {White}, {Habib},
  {Williams}, {Lawrence}, \& {Wagner}}]{Heitmann2009}
{Heitmann}, K., {Higdon}, D., {White}, M., {et~al.} 2009,
  \href{http://dx.doi.org/10.1088/0004-637X/705/1/156}{\JournalTitle{\apj},
  705, 156}

\bibitem[{{Heitmann} {et~al.}(2016){Heitmann}, {Bingham}, {Lawrence},
  {Bergner}, {Habib}, {Higdon}, {Pope}, {Biswas}, {Finkel}, {Frontiere}, \&
  {Bhattacharya}}]{Heitmann2016}
{Heitmann}, K., {Bingham}, D., {Lawrence}, E., {et~al.} 2016,
  \href{http://dx.doi.org/10.3847/0004-637X/820/2/108}{\JournalTitle{\apj},
  820, 108}

\bibitem[{{Hinshaw} {et~al.}(2013){Hinshaw}, {Larson}, {Komatsu}, {Spergel},
  {Bennett}, {Dunkley}, {Nolta}, {Halpern}, {Hill}, {Odegard}, {Page}, {Smith},
  {Weiland}, {Gold}, {Jarosik}, {Kogut}, {Limon}, {Meyer}, {Tucker}, {Wollack},
  \& {Wright}}]{WMAP9}
{Hinshaw}, G., {Larson}, D., {Komatsu}, E., {et~al.} 2013,
  \href{http://dx.doi.org/10.1088/0067-0049/208/2/19}{\JournalTitle{\apjs},
  208, 19}

\bibitem[{Hunter(2007)}]{matplotlib}
Hunter, J.~D. 2007,
  \href{http://dx.doi.org/10.1109/MCSE.2007.55}{\JournalTitle{Computing in
  Science Engineering}, 9, 90}

\bibitem[{{Jenkins} {et~al.}(2001){Jenkins}, {Frenk}, {White}, {Colberg},
  {Cole}, {Evrard}, {Couchman}, \& {Yoshida}}]{Jenkins01}
{Jenkins}, A., {Frenk}, C.~S., {White}, S.~D.~M., {et~al.} 2001,
  \JournalTitle{\mnras}, 321, 372

\bibitem[{Jones {et~al.}(2001--)Jones, Oliphant, Peterson, {et~al.}}]{scipy}
Jones, E., Oliphant, T., Peterson, P., {et~al.} 2001--, {SciPy}: Open source
  scientific tools for {Python}, [Online;
  \href{http://www.scipy.org/}{scipy.org}]

\bibitem[{{Lesgourgues}(2011)}]{CLASS1}
{Lesgourgues}, J. 2011, \JournalTitle{ArXiv e-prints},
  \href{http://arxiv.org/abs/1104.2932}{{\sffamily arXiv:1104.2932
  [astro-ph.IM]}}

\bibitem[{Lewis \& Bridle(2002)}]{CAMB}
Lewis, A., \& Bridle, S. 2002,
  \href{http://dx.doi.org/10.1103/PhysRevD.66.103511}{\JournalTitle{Phys.
  Rev.}, D66, 103511}

\bibitem[{{Mantz} {et~al.}(2016){Mantz}, {Allen}, {Morris}, {von der Linden},
  {Applegate}, {Kelly}, {Burke}, {Donovan}, \& {Ebeling}}]{Mantz2016}
{Mantz}, A.~B., {Allen}, S.~W., {Morris}, R.~G., {et~al.} 2016,
  \href{http://dx.doi.org/10.1093/mnras/stw2250}{\JournalTitle{\mnras}, 463,
  3582}

\bibitem[{{McClintock} {et~al.}(2018){McClintock}, {Varga}, {Melchior}, \&
  {Gruen}}]{McClintock2018}
{McClintock}, T., {Varga}, T.~N., {Melchior}, P., \& {Gruen}, D. 2018,
  \JournalTitle{ArXiv e-prints}

\bibitem[{{Mehrtens} {et~al.}(2012){Mehrtens}, {Romer}, {Hilton},
  {Lloyd-Davies}, {Miller}, {Stanford}, {Hosmer}, {Hoyle}, {Collins}, {Liddle},
  {Viana}, {Nichol}, {Stott}, {Dubois}, {Kay}, {Sahl{\'e}n}, {Young}, {Short},
  {Christodoulou}, {Watson}, {Davidson}, {Harrison}, {Baruah}, {Smith},
  {Burke}, {Mayers}, {Deadman}, {Rooney}, {Edmondson}, {West}, {Campbell},
  {Edge}, {Mann}, {Sabirli}, {Wake}, {Benoist}, {da Costa}, {Maia}, \&
  {Ogando}}]{Mehrtens2012}
{Mehrtens}, N., {Romer}, A.~K., {Hilton}, M., {et~al.} 2012,
  \href{http://dx.doi.org/10.1111/j.1365-2966.2012.20931.x}{\JournalTitle{\mnras},
  423, 1024}

\bibitem[{{Melchior} {et~al.}(2017){Melchior}, {Gruen}, {McClintock}, {Varga},
  {Sheldon}, {Rozo}, {Amara}, {Becker}, {Benson}, {Bermeo}, {Bridle},
  {Clampitt}, {Dietrich}, {Hartley}, {Hollowood}, {Jain}, {Jarvis}, {Jeltema},
  {Kacprzak}, {MacCrann}, {Rykoff}, {Saro}, {Suchyta}, {Troxel}, {Zuntz},
  {Bonnett}, {Plazas}, {Abbott}, {Abdalla}, {Annis}, {Benoit-L{\'e}vy},
  {Bernstein}, {Bertin}, {Brooks}, {Buckley-Geer}, {Carnero Rosell}, {Carrasco
  Kind}, {Carretero}, {Cunha}, {D'Andrea}, {da Costa}, {Desai}, {Eifler},
  {Flaugher}, {Fosalba}, {Garc{\'{\i}}a-Bellido}, {Gaztanaga}, {Gerdes},
  {Gruendl}, {Gschwend}, {Gutierrez}, {Honscheid}, {James}, {Kirk}, {Krause},
  {Kuehn}, {Kuropatkin}, {Lahav}, {Lima}, {Maia}, {March}, {Martini},
  {Menanteau}, {Miller}, {Miquel}, {Mohr}, {Nichol}, {Ogando}, {Romer},
  {Sanchez}, {Scarpine}, {Sevilla-Noarbe}, {Smith}, {Soares-Santos},
  {Sobreira}, {Swanson}, {Tarle}, {Thomas}, {Walker}, {Weller}, \&
  {Zhang}}]{Melchior2017}
{Melchior}, P., {Gruen}, D., {McClintock}, T., {et~al.} 2017,
  \href{http://dx.doi.org/10.1093/mnras/stx1053}{\JournalTitle{\mnras}, 469,
  4899}

\bibitem[{{Miyazaki} {et~al.}(2012)}]{HSC12}
{Miyazaki}, S., {et~al.} 2012, \href{http://dx.doi.org/10.1117/12.926844}{in
  Society of Photo-Optical Instrumentation Engineers (SPIE) Conference Series,
  Vol. 8446, Society of Photo-Optical Instrumentation Engineers (SPIE)
  Conference Series}

\bibitem[{{More} {et~al.}(2011){More}, {Kravtsov}, {Dalal}, \&
  {Gottl{\"o}ber}}]{More2011}
{More}, S., {Kravtsov}, A.~V., {Dalal}, N., \& {Gottl{\"o}ber}, S. 2011,
  \href{http://dx.doi.org/10.1088/0067-0049/195/1/4}{\JournalTitle{\apjs}, 195,
  4}

\bibitem[{{Piffaretti} {et~al.}(2011){Piffaretti}, {Arnaud}, {Pratt},
  {Pointecouteau}, \& {Melin}}]{Piffaretti2011}
{Piffaretti}, R., {Arnaud}, M., {Pratt}, G.~W., {Pointecouteau}, E., \&
  {Melin}, J.-B. 2011,
  \href{http://dx.doi.org/10.1051/0004-6361/201015377}{\JournalTitle{\aap},
  534, A109}

\bibitem[{{Planck Collaboration} {et~al.}(2014){Planck Collaboration}, {Ade},
  {Aghanim}, {Armitage-Caplan}, {Arnaud}, {Ashdown}, {Atrio-Barandela},
  {Aumont}, {Baccigalupi}, {Banday}, \& et~al.}]{PlanckXVI}
{Planck Collaboration}, {Ade}, P.~A.~R., {Aghanim}, N., {et~al.} 2014,
  \href{http://dx.doi.org/10.1051/0004-6361/201321591}{\JournalTitle{\aap},
  571, A16}

\bibitem[{{Planck Collaboration} {et~al.}(2016){Planck Collaboration}, {Ade},
  {Aghanim}, {Arnaud}, {Ashdown}, {Aumont}, {Baccigalupi}, {Banday},
  {Barreiro}, {Bartlett}, \& et~al.}]{PlanckXXIV2015}
---. 2016,
  \href{http://dx.doi.org/10.1051/0004-6361/201525833}{\JournalTitle{\aap},
  594, A24}

\bibitem[{{Press} \& {Schechter}(1974)}]{PressSchechter74}
{Press}, W.~H., \& {Schechter}, P. 1974, \JournalTitle{\apj}, 187, 425

\bibitem[{{Radovich} {et~al.}(2017){Radovich}, {Puddu}, {Bellagamba},
  {Roncarelli}, {Moscardini}, {Bardelli}, {Grado}, {Getman}, {Maturi}, {Huang},
  {Napolitano}, {McFarland}, {Valentijn}, \& {Bilicki}}]{Radovich2017}
{Radovich}, M., {Puddu}, E., {Bellagamba}, F., {et~al.} 2017,
  \href{http://dx.doi.org/10.1051/0004-6361/201629353}{\JournalTitle{\aap},
  598, A107}

\bibitem[{Rasmussen \& Williams(2005)}]{RasmussenWilliamsGPs}
Rasmussen, C.~E., \& Williams, C. K.~I. 2005, Gaussian Processes for Machine
  Learning (Adaptive Computation and Machine Learning) (The MIT Press)

\bibitem[{{Reed} {et~al.}(2007){Reed}, {Bower}, {Frenk}, {Jenkins}, \&
  {Theuns}}]{Reed2007}
{Reed}, D.~S., {Bower}, R., {Frenk}, C.~S., {Jenkins}, A., \& {Theuns}, T.
  2007,
  \href{http://dx.doi.org/10.1111/j.1365-2966.2006.11204.x}{\JournalTitle{\mnras},
  374, 2}

\bibitem[{{Reed} {et~al.}(2013){Reed}, {Smith}, {Potter}, {Schneider},
  {Stadel}, \& {Moore}}]{Reed2013}
{Reed}, D.~S., {Smith}, R.~E., {Potter}, D., {et~al.} 2013,
  \href{http://dx.doi.org/10.1093/mnras/stt301}{\JournalTitle{\mnras}, 431,
  1866}

\bibitem[{{Rykoff} {et~al.}(2014){Rykoff}, {Rozo}, {Busha}, {Cunha},
  {Finoguenov}, {Evrard}, {Hao}, {Koester}, {Leauthaud}, {Nord}, {Pierre},
  {Reddick}, {Sadibekova}, {Sheldon}, \& {Wechsler}}]{Rykoff2014}
{Rykoff}, E.~S., {Rozo}, E., {Busha}, M.~T., {et~al.} 2014,
  \href{http://dx.doi.org/10.1088/0004-637X/785/2/104}{\JournalTitle{\apj},
  785, 104}

\bibitem[{{Rykoff} {et~al.}(2016){Rykoff}, {Rozo}, {Hollowood},
  {Bermeo-Hernandez}, {Jeltema}, {Mayers}, {Romer}, {Rooney}, {Saro}, {Vergara
  Cervantes}, {Wechsler}, {Wilcox}, {Abbott}, {Abdalla}, {Allam}, {Annis},
  {Benoit-L{\'e}vy}, {Bernstein}, {Bertin}, {Brooks}, {Burke}, {Capozzi},
  {Carnero Rosell}, {Carrasco Kind}, {Castander}, {Childress}, {Collins},
  {Cunha}, {D'Andrea}, {da Costa}, {Davis}, {Desai}, {Diehl}, {Dietrich},
  {Doel}, {Evrard}, {Finley}, {Flaugher}, {Fosalba}, {Frieman}, {Glazebrook},
  {Goldstein}, {Gruen}, {Gruendl}, {Gutierrez}, {Hilton}, {Honscheid}, {Hoyle},
  {James}, {Kay}, {Kuehn}, {Kuropatkin}, {Lahav}, {Lewis}, {Lidman}, {Lima},
  {Maia}, {Mann}, {Marshall}, {Martini}, {Melchior}, {Miller}, {Miquel},
  {Mohr}, {Nichol}, {Nord}, {Ogando}, {Plazas}, {Reil}, {Sahl{\'e}n},
  {Sanchez}, {Santiago}, {Scarpine}, {Schubnell}, {Sevilla-Noarbe}, {Smith},
  {Soares-Santos}, {Sobreira}, {Stott}, {Suchyta}, {Swanson}, {Tarle},
  {Thomas}, {Tucker}, {Uddin}, {Viana}, {Vikram}, {Walker}, {Zhang}, \& {DES
  Collaboration}}]{Rykoff2016}
{Rykoff}, E.~S., {Rozo}, E., {Hollowood}, D., {et~al.} 2016,
  \href{http://dx.doi.org/10.3847/0067-0049/224/1/1}{\JournalTitle{\apjs}, 224,
  1}

\bibitem[{{Sheth} {et~al.}(2001){Sheth}, {Mo}, \& {Tormen}}]{ShethMoTormen01}
{Sheth}, R.~K., {Mo}, H.~J., \& {Tormen}, G. 2001, \JournalTitle{\mnras}, 323,
  1

\bibitem[{{Sheth} \& {Tormen}(1999)}]{ShethTormen99}
{Sheth}, R.~K., \& {Tormen}, G. 1999, \JournalTitle{\mnras}, 308, 119

\bibitem[{{Simet} {et~al.}(2017){Simet}, {McClintock}, {Mandelbaum}, {Rozo},
  {Rykoff}, {Sheldon}, \& {Wechsler}}]{Simet2017}
{Simet}, M., {McClintock}, T., {Mandelbaum}, R., {et~al.} 2017,
  \href{http://dx.doi.org/10.1093/mnras/stw3250}{\JournalTitle{\mnras}, 466,
  3103}

\bibitem[{{Springel}(2005)}]{Springel05}
{Springel}, V. 2005,
  \href{http://dx.doi.org/10.1111/j.1365-2966.2005.09655.x}{\JournalTitle{\mnras},
  364, 1105}

\bibitem[{{Stern} {et~al.}(2018){Stern}, {Dietrich}, {Bocquet}, {Applegate},
  {Mohr}, {Bridle}, {Carrasco Kind}, {Gruen}, {Jarvis}, {Kacprzak}, {Saro},
  {Sheldon}, {Troxel}, {Zuntz}, {Benson}, {Capasso}, {Chiu}, {Desai},
  {Rapetti}, {Reichardt}, {Saliwanchik}, {Schrabback}, {Gupta}, {Abbott},
  {Abdalla}, {Avila}, {Bertin}, {Brooks}, {Burke}, {Carnero Rosell},
  {Carretero}, {Castander}, {D'Andrea}, {da Costa}, {Davis}, {De Vicente},
  {Diehl}, {Doel}, {Estrada}, {Evrard}, {Flaugher}, {Fosalba}, {Frieman},
  {Garc{\'{\i}}a-Bellido}, {Gaztanaga}, {Gruendl}, {Gschwend}, {Gutierrez},
  {Hollowood}, {Jeltema}, {Kirk}, {Kuehn}, {Kuropatkin}, {Lahav}, {Lima},
  {Maia}, {March}, {Melchior}, {Menanteau}, {Miquel}, {Plazas}, {Romer},
  {Sanchez}, {Schindler}, {Schubnell}, {Sevilla-Noarbe}, {Smith}, {Smith},
  {Sobreira}, {Suchyta}, {Swanson}, {Tarle}, \& {Walker}}]{Stern2018}
{Stern}, C., {Dietrich}, J.~P., {Bocquet}, S., {et~al.} 2018,
  \JournalTitle{ArXiv e-prints},
  \href{http://arxiv.org/abs/1802.04533}{{\sffamily arXiv:1802.04533}}

\bibitem[{{Suzuki} {et~al.}(2012){Suzuki}, {Rubin}, {Lidman}, {Aldering},
  {Amanullah}, {Barbary}, {Barrientos}, {Botyanszki}, {Brodwin}, {Connolly},
  {Dawson}, {Dey}, {Doi}, {Donahue}, {Deustua}, {Eisenhardt}, {Ellingson},
  {Faccioli}, {Fadeyev}, {Fakhouri}, {Fruchter}, {Gilbank}, {Gladders},
  {Goldhaber}, {Gonzalez}, {Goobar}, {Gude}, {Hattori}, {Hoekstra}, {Hsiao},
  {Huang}, {Ihara}, {Jee}, {Johnston}, {Kashikawa}, {Koester}, {Konishi},
  {Kowalski}, {Linder}, {Lubin}, {Melbourne}, {Meyers}, {Morokuma}, {Munshi},
  {Mullis}, {Oda}, {Panagia}, {Perlmutter}, {Postman}, {Pritchard}, {Rhodes},
  {Ripoche}, {Rosati}, {Schlegel}, {Spadafora}, {Stanford}, {Stanishev},
  {Stern}, {Strovink}, {Takanashi}, {Tokita}, {Wagner}, {Wang}, {Yasuda},
  {Yee}, \& {Supernova Cosmology Project}}]{Suzuki2012}
{Suzuki}, N., {Rubin}, D., {Lidman}, C., {et~al.} 2012,
  \href{http://dx.doi.org/10.1088/0004-637X/746/1/85}{\JournalTitle{\apj}, 746,
  85}

\bibitem[{{Tinker} {et~al.}(2008){Tinker}, {Kravtsov}, {Klypin}, {Abazajian},
  {Warren}, {Yepes}, {Gottl{\"o}ber}, \& {Holz}}]{Tinker2008}
{Tinker}, J., {Kravtsov}, A.~V., {Klypin}, A., {et~al.} 2008,
  \href{http://dx.doi.org/10.1086/591439}{\JournalTitle{\apj}, 688, 709}

\bibitem[{{Tinker} {et~al.}(2010){Tinker}, {Robertson}, {Kravtsov}, {Klypin},
  {Warren}, {Yepes}, \& {Gottl{\"o}ber}}]{Tinker10}
{Tinker}, J.~L., {Robertson}, B.~E., {Kravtsov}, A.~V., {et~al.} 2010,
  \href{http://dx.doi.org/10.1088/0004-637X/724/2/878}{\JournalTitle{\apj},
  724, 878}

\bibitem[{van~der Walt {et~al.}(2011)van~der Walt, Colbert, \&
  Varoquaux}]{numpy}
van~der Walt, S., Colbert, S.~C., \& Varoquaux, G. 2011,
  \href{http://dx.doi.org/10.1109/MCSE.2011.37}{\JournalTitle{Computing in
  Science Engineering}, 13, 22}

\bibitem[{{Warren} {et~al.}(2006){Warren}, {Abazajian}, {Holz}, \&
  {Teodoro}}]{Warren06}
{Warren}, M.~S., {Abazajian}, K., {Holz}, D.~E., \& {Teodoro}, L. 2006,
  \href{http://dx.doi.org/10.1086/504962}{\JournalTitle{\apj}, 646, 881}

\bibitem[{{Weinberg} {et~al.}(2013){Weinberg}, {Mortonson}, {Eisenstein},
  {Hirata}, {Riess}, \& {Rozo}}]{Weinberg13}
{Weinberg}, D.~H., {Mortonson}, M.~J., {Eisenstein}, D.~J., {et~al.} 2013,
  \href{http://dx.doi.org/10.1016/j.physrep.2013.05.001}{\JournalTitle{\physrep},
  530, 87}

\bibitem[{{Zu} {et~al.}(2014){Zu}, {Weinberg}, {Rozo}, {Sheldon}, {Tinker}, \&
  {Becker}}]{Zu14}
{Zu}, Y., {Weinberg}, D.~H., {Rozo}, E., {et~al.} 2014,
  \href{http://dx.doi.org/10.1093/mnras/stu033}{\JournalTitle{\mnras}, 439,
  1628}

\end{thebibliography}

\appendix


\section{Assessing the emulator uncertainty model}
\label{app:verifying_em_unc_model}

\begin{figure*}[htb!]
	\label{fig:residual_histograms}
    \includegraphics[width=\linewidth]{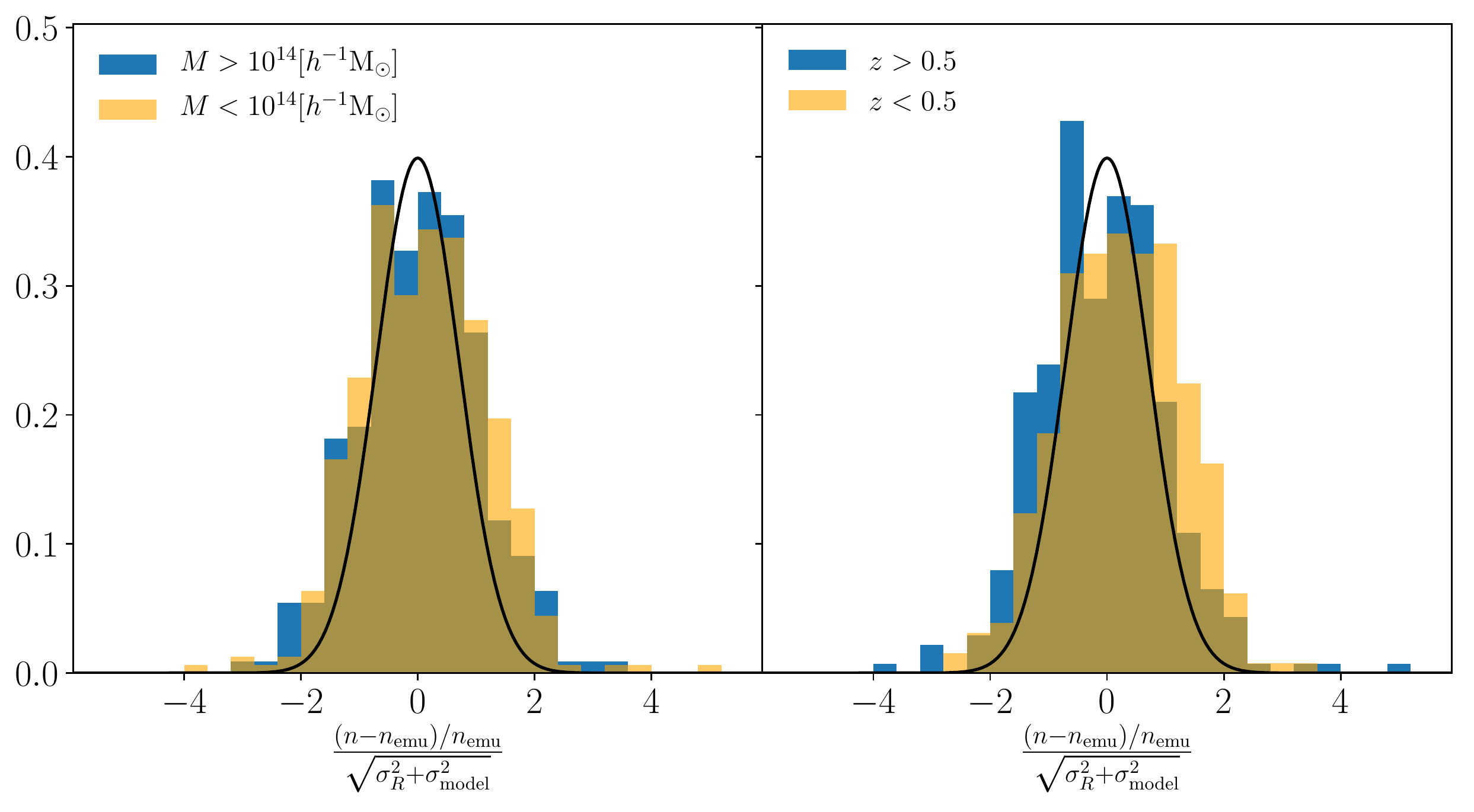}
    \caption{Histograms of the residuals of all snapshots for all testing simulations divided by the square root of their variances plus the emulator uncertainties added in quadrature, as compared to a unit normal distribution in black. {\it Left}: Residuals split by mass at $10^{14}\ \hmsun$. {\it Right}: Residuals split by redshift at $z=0.5$.  We caution that this figure cannot be used to gauge whether the calibration of the emulator uncertainty is correct via $\chi$-by-eye, as neighboring mass and redshift bins are correlated.  A goodness-of-fit test that accounts for these covariances results in acceptable $\chi^2$ values.}
\end{figure*}

Residuals between the emulator and the simulations occur for three reasons: systematic differences between the emulator and the true mass function, Poisson noise in the number of halos, and sample variance due to the particular realization of density modes present in the simulation. Poisson noise and sample variance are estimated through the jackknife covariance matrix from each simulation snapshot. The accuracy of the emulator $\sigma_{\rm model}$ is calibrated as described in detail in \autoref{sec:modeling_the_residuals}. To validate the estimated accuracy of the emulator, we define
\begin{equation}
	\chi_R = \frac{(n-n_{\rm emu})/n_{\rm emu}}{\sqrt{\sigma_R^2+\sigma_{\rm model}^2}}\,.
\end{equation}
\autoref{fig:residual_histograms} shows the distribution of $\chi_R$ for two splits of the residuals, by mass and by redshift, for all snapshots for bins with more than 20 halos. The splits were chosen to have approximately the same number of residuals on either side. Outliers ($>3\sigma$) occurred as follows: one out of 43 bins at $z=3$, three out of 44 bins at $z=2$, and one each out of 83 and 85 bins at $z=0.1$ and $z=0$, respectively. This is six mass bins out of a total of 668 across all redshifts in all simulations. The \citet{Tinker2008} model was significantly more discrepant at all redshifts compared to the emulator. 

While \autoref{fig:residual_histograms} is useful for visualization, the fact that different bins are correlated implies that we cannot use the distribution of residuals $\chi_R$ to establish goodness of fit. Instead, we compute the total $\chi^2$ of the emulator across all simulations for each individual snapshot. For each individual snapshot of each simulation, we computed
\begin{equation}
	\label{eq:chi2_for_resids}
    \chi^2 = {\bf R}\widetilde{\bf C}^{-1}{\bf R}^T\,,
\end{equation}
where $\widetilde{\bf C}$ is defined in \autoref{eq:residual_full_covariance}, and we incorporate the finite resolution correction discussed in \autoref{sec:halo_identification} as additional uncertainty. We then summed over all our test simulations to arrive at a final $\chi^2$.  
We found all snapshots exhibited an acceptable total $\chi^2$ across all simulations for $z\leq 2$.
At $z=3$, the $\chi^2$ of the emulator is slightly high, yielding $\chi^2/dof = 47/28$. The worst of the remaining $\chi^2$ values is $\chi^2/dof=96/85$ at $z=0$. These results demonstrate that the emulator accurately predicts the mass function up to at least $z=2$.  At $z=3$, the emulator might be slightly over-optimistic, though the difference is clearly not strongly significant.

\section{Accuracy Requirements}
\label{app:accuracy_requirements}

The required accuracy of the halo mass function is set by considering the uncertainty on cluster abundances in real surveys. In a survey, clusters are binned by some observable that is related to mass, for example \redmapper\ richness \citep{Rykoff2014}, X-ray luminosity or temperature, or Sunyaev-Zeldovich signal.  The mean cluster mass of objects in these bins is measured using techniques such as weak lensing. If clusters are binned by mass, the number of clusters in a single bin is
\begin{equation}
	\label{eq:n_clust_in_bin}
    \ln N = \Delta\ln M \frac{{\rm d}\ln N}{{\rm d}\ln M}\,,
\end{equation}
where we assume a fiducial bin width of $\Delta\ln M = 0.2\ln 10$, or five bins per decade. If the bin has a true mean mass $\ln M_b$ measured with some uncertainty $\sigma_{\ln M}$, then error in the abundance is found via propagation of errors.  One has
\begin{eqnarray}
	\label{eq:propagating_mass_uncertainty}
    \ln N(\ln M_b + \sigma_{\ln M}) &=& \Delta\ln M \frac{{\rm d}\ln N}{{\rm d}\ln M}\bigg\rvert_{\ln M_b+\sigma_{\ln M}} \\
    &=& \Delta\ln M \left(\frac{{\rm d}\ln N}{{\rm d}\ln M}\bigg\rvert_{\ln M_b} + \sigma_{\ln M} \frac{{\rm d}^2\ln N}{{\rm d}\ln M^2}\bigg\rvert_{\ln M_b}  \right) \\
    &=& \ln N + \Delta\ln M \sigma_{\ln M} \frac{{\rm d}^2\ln N}{{\rm d}\ln M^2}\bigg\rvert_{\ln M_b}\,.
\end{eqnarray}
Note that ${\rm d}^2 \ln N/{\rm d}\ln M^2$ is the first derivative of the halo mass function.
The uncertainty on the abundance due to uncertainty in the mass is %
\begin{equation}
	\sigma_{\ln N, M} = \Delta\ln M \sigma_{\ln M} \frac{{\rm d}^2 \ln N}{{\rm d}\ln M^2}\bigg\rvert_{\ln M_b}\,.
\end{equation}
\citet{McClintock2018} found the statistical uncertainty on the mean masses of clusters in the DES Y1 data to be $\sigma_{\ln M} \sim 8\%$. Of course, this uncertainty will change between surveys and is sensitive to lens source density, survey area, and survey depth. We consider two contributions to the uncertainty in weak lensing mass calibration: shape noise (the dominant source of statistical uncertainty for the DES), and cluster sample variance
\begin{equation}
	\label{eq:sigma_lnM}
    \sigma_{\ln M}^2 = \sigma_{\ln M}^{2, {\rm SN}} + \sigma_{\ln M}^{2, {\rm SV}}\,.
\end{equation}
The sample variance is easily understood:
\begin{equation}
	\label{eq:sigma_lnM_SV}
    \sigma_{\ln M}^{2, {\rm SV}} = \frac{{\rm Var}(M|\lambda)}{N} = \frac{0.4^2}{N}\,,
\end{equation}
where the scatter in mass at fixed richness is roughly 40\%. This term is subdominant to the shape noise, which is more complicated. \citet{Melchior2017} found that $M \propto \Delta\Sigma^{4/3}$ where $\Delta\Sigma$ is the weak lensing profile. This means that the uncertainty in the mass due to shape noise is given by
\begin{equation}
	\label{eq:sigma_M_SN}
    \sigma_M^{2, {\rm SN}} =\sigma^2_{\Delta\Sigma}M^{1/2} = A M^{1/2}(N_cn)^{-1}\,,
\end{equation}
where $N_c$ is the number of clusters, $A$ is some constant, and $n$ is the lensing source density. However, since clusters at higher redshift have fewer sources behind them, this term is redshift dependent. In detail this dependence is complicated, but it will generally resemble
\begin{equation}
	n(z) \approx n_0 \exp\left(-\frac{1}{2}\frac{z^2}{z_*^2} \right)\,,
\end{equation}
where $z_*$ is a characteristic redshift where the local source density peaks. For DES Y1, $n_0$ is 6.3 arcmin$^{-2}$. We found that the parameters $A=0.7\times 10^{22}\ [\hmsun]^{1.5}\ {\rm arcmin^{-2}}$ and $z_*=0.33$ recovered the uncertainties of the DES Y1 masses within 10\%. In a generic mass bin at mass $M$ for a given survey, we can use these values to compute the shape noise uncertainty using
\begin{equation}
	\sigma_M^{2, {\rm SN}} = \frac{AM^{1/2}}{N}\frac{6.3\ {\rm arcmin^{-2}}}{n_0} \exp\left(-\frac{1}{2}\frac{z^2}{z_*^2} \right)\,.
\end{equation}
The $z_*$ values for DES Y1, Y5 and LSST Y1 are $0.5, 0.6$, and $8.5$ respectively. The number of clusters in the bin $N$ scales with the survey area, and is calculated within a redshift slice of $\Delta z=0.15$, the width in redshift of the cluster bins in \citet{McClintock2018}. The lens source densities for DES Y5 and LSST Y1 are estimated to be $8.4$ and 10 arcmin$^{-2}$. The survey areas for DES Y1, Y5, and LSST Y1 are 1514, 5000, and 18000 deg$^2$. The shape noise uncertainty in $\ln M$ is $\sigma_{\ln M}^{\rm SN} = \sigma_M^{\rm SN}/M$, which allows us to compute the total uncertainty on the mass bin $\sigma_{\ln M}$. From this we calculate the total uncertainty on the abundance as
\begin{equation}
	\sigma_{\ln N}^2 = \sigma_{\ln N, {\rm emu}}^2 + \sigma_{\ln N, M}^2\,,
\end{equation}
where $\sigma_{\ln N, {\rm emu}}$ is the accuracy of the halo mass function emulator. We require that adding the emulator uncertainty increases the error budget for each survey by no more than 10\%, so that $\sigma_{\ln N}^2 = 1.1^2\sigma_{\ln N, M}^2$.  We use this equation to derive the calibration requirements shown in \autoref{fig:accuracy_requirements}.

\end{document}